\documentclass[english,aps,prd,superscriptaddress,twocolumn]{revtex4-1}
\usepackage[T1]{fontenc}
\usepackage[latin9]{inputenc}
\usepackage{amsmath}
\usepackage{graphicx}

\makeatletter
\usepackage{slashed}
\usepackage{hyperref}
\usepackage{MnSymbol}
\usepackage{bbm}

\makeatother

\usepackage{babel}
\begin{document}
\title{Torsional Anomalies and Bulk-Dislocation Correspondence in Weyl Systems}
\author{Ze-Min Huang }
\email{zeminh2@illinois.edu}

\affiliation{Department of Physics, University of Illinois at Urbana-Champaign,
1110 West Green Street, Urbana, Illinois 61801, USA}
\author{Bo Han}
\affiliation{Theory of Condensed Matter Group, Cavendish Laboratory, University
of Cambridge, J. J. Thomson Avenue, Cambridge CB3 0HE, United Kingdom}
\date{\today}
\begin{abstract}
Based on the supersymmetric quantum mechanical approach, we have systematically
studied both the $U\left(1\right)$ gauge anomaly and the diffeomorphism
anomaly in Weyl systems with torsion, curvature and external electromagnetic
fields. These anomalies relate to the chiral current (or current)
non-conservation and chiral energy-momentum (or energy-momentum) non-conservation,
respectively, which can be applied to the $^{3}\text{He-A}$ phase,
the chiral superconductors and the Weyl semimetals with dislocations
and disclinations. In sharp difference with other anomalies, there
exist torsional anomalies depending on the position of Weyl nodes
in the energy-momentum space. These anomalies originate from particles
pumped up through the Weyl nodes and they are thus insensitive to
the ultra-violet physics, while the Nieh-Yan anomaly is from the particle
inflow through the ultra-violet cut-off. The current non-conservation
as well as the energy-momentum non-conservation are found, which stem
from the zero modes trapped in the dislocations and they can be understood
from the Callan-Harvey mechanism. Finally, by comparing our results
with the well-established momentum anomaly in the  $^{3}\text{He-A}$
phase, the Nieh-Yan term as well as other cut-off dependent terms
are shown to be negligible, because the ratio between the Lorentz
symmetry breaking scale and the chemical potential is of order $10^{-5}$. 
\end{abstract}
\maketitle

\section{Introduction}

Torsion can induce chiral current non-conservation, which is known
as the Nieh-Yan anomaly \citep{nieh1982aop,nieh1982jmp,yajima1985ptp,chandia1997prd,obukhov1997fop,peeters1999jhep,parrikar2014prd}.
However, the Nieh-Yan anomaly depends on the ultra-violet cut-off
and thus the ultra-violet physics, so it is in controversial since
its discovery \citep{kreimer2001prd,chandia2001prd}. In addition
to this, a series of torsional anomalies in various dimensions are
also found \citep{hughes2011prl,parrikar2014prd,huang2019arxiv_ham},
including the diffeomorphism anomaly and the parity anomaly. They
all depend on the ultra-violet cut-off. In condensed matter physics,
torsion naturally arises from the pairing order parameters in the
$^{3}\text{He-A}$ phase \citep{leggett1975rmp,volovik1992ws,volovik2003oxford,golan2018prb,Nissinen2019arxivNY}
and the chiral superconductors \citep{Kallin2016iop,koziie2016sa,Ran2019science,jiao2019arxiv,metz2019prb},
dislocations \citep{katanaev1992aop,zubkov2015aop,sumiyoshi2016prl,corjito2016aop,huang2018prb},
back ground rotations and temperature gradient \citep{shitade2014ptep,tatara2015prl,huang2019arxiv_thermal}.
Many novel topological phenomena in Weyl/Dirac systems have close
connection with quantum anomalies \citep{stone1985prl,balatskii1986zetf,volovik1986jetp,stone1987aop,landsteiner2011prl,zyuzin2012prb,son2013prb,pikulin2016prx,grushin2016prx,huang2017prb,stone2018prd,burkov2018arcmp,armitage2018rmp},
including the momentum anomaly \citep{cross1975jltp,mermin1980prb,stone1985prl,volovik1986jetp,combescot1986prb,balatskii1986zetf,stone1987aop,bevan1997nature},
the anomalous quantum Hall effect \citep{yang2011prb,zyuzin2012prb},
the chiral magnetic effect \citep{vilenkin1980prd,fukushima2008prd,li2015naturecom,li2016naturecom,li2016naturephy}
and the negative magneto-thermoelectric resistance \citep{gooth2017nature}.
Recently, torsional anomalies are attracting much attention. For example,
they have been used to describe various viscoelastic responses \citep{hughes2011prl,hughes2013prd,parrikar2014prd,sun2014epl,shapourian2015prb,you2016prb,sumiyoshi2016prl,huang2018prb,nissinen2019prr}
and the anomalous thermal Hall effect \citep{huang2019arxiv_thermal}.
However, due to the cut-off dependence in the torsional anomalies,
one might question about the subtle role of the Nieh-Yan anomaly.
In contrast to the high-energy physics, the ultra-violet physics is
known in condensed matters. So in principle, one can fix these ambiguities
in the Nieh-Yan anomaly and other cut-off dependent anomalies. 

In Weyl systems, there are gapless nodes protected by the monopole
charge in the momentum space. Due to the time-reversal symmetry or
inversion symmetry breaking, Weyl nodes with different chiralities
locate at different positions in the energy-momentum space, which
provide a new scale. The anomalies from the gauge fields and the gravitational
fields have close connection with the Atiyah-Singer index theorem
and they do not depend on the energy/momentum scale, while the torsional
anomalies do. Hence, it is natural to ask if the scale from Weyl nodes'
position in the energy-momentum space can enter the anomaly equations
and lead to new anomalies. 

It is known that there is momentum anomaly in the $^{3}\text{He-A}$
phase, the momentum current of the excitations are not conserved \citep{combescot1986prb,stone1985prl,volovik1985jetp_normal,balatskii1986zetf,volovik1986jetp,volovik1986jetp_gravity,stone1987aop,bevan1997nature}.
The corresponding expermental observation was reported in Ref. \citep{bevan1997nature}.
Theoretically, the authors of Ref. \citep{stone1985prl,stone1987aop}
have noticed that the Fermi surface can effectively reduce the system
dimensions from $4$ to $2$. Then the Wilczek-Goldstone formula \citep{goldstone1981prl}
is applicable. By focusing on the region away from the gapless nodes,
they have successfully derived the momentum anomaly from the $\left(1+1\right)$-dim
chiral anomaly. By contrast, in Ref. \citep{balatskii1986zetf,volovik1986jetp,volovik1986jetp_gravity},
the authors have noticed that around the gapless nodes, the texture
behaves like the axial gauge fields, which enables them to understand
the momentum anomaly in terms of the $\left(3+1\right)$-dim chiral
anomaly. Since the direct derivation of the momentum anomaly is still
absent, we are motivated to derive the corresponding diffeomorphism
anomaly directly from the quantum field theory, rather than resorting
to the chiral anomaly. 

In this paper, by using the method of supersymmetric quantum mechanics
\citep{alvarez1983cmp,friedan1985pd,boer1996npb,bastianelli2006cambridge},
we have derived both the (chiral) $U\left(1\right)$ gauge anomaly
and the (chiral) diffeomorphism anomaly for the $^{3}\text{He-A}$
phase, the chiral superconductors and the Weyl semimetals with dislocations
(see Eq. (\ref{eq:u(1)_anomaly}) and Eq. (\ref{eq:diffeomorphism_anomaly})).
Compared to terms known for Dirac Fermions \citep{yajima1985ptp,peeters1999jhep,parrikar2014prd,chandia1997prd,chandia2001prd,obukhov1997fop},
we have found that the scale given by the displacement of Weyl Fermions
in the energy-momentum space can lead to new anomaly terms. For example,
the conservation law for both the $U\left(1\right)$ current and the
energy-momentum current is superficially violated, which is because
of the chiral zero modes trapped in the dislocations. The chirality
of the zero modes is determined by the sign of the product of the
Burgers vector and the separation between Weyl nodes. This is consistent
with the criterion in topological insulators \citep{ran2009np,yuji2013prl,taylor2014prb},
for example, by setting the separation between Weyl nodes to the reciprocal
vector. Especially, by imposing the particle-hole symmetry, these
becomes the chiral Majorana zero modes. We have recovered the momentum
anomaly as well. In addition to the term obtained in Ref. \citep{stone1985prl,combescot1986prb,balatskii1986zetf,volovik1986jetp,stone1987aop},
we have found terms depending on the cut-off. In the  $^{3}\text{He-A}$
phase, the ratio between the Lorentz symmetry breaking scale and the
chemical potential (or equivalently, the scale from the position of
Weyl nodes in the energy-momentum space) is $10^{-5}$ \citep{volovik2003oxford},
so the Nieh-Yan term as well as other cut-off dependent terms in the
momentum anomaly are shown to be negligible. In summary, the main
results of this paper are threefold. Firstly, new anomaly terms are
derived for the first time (see Eq. (\ref{eq:energy_momentum_delta})),
from which, the bulk-dislocation correspondence are constructed. Especially,
based on the diffeomorphism anomaly, we have found chiral Majorana
zero modes trapped in the dislocations in chiral superconductors.
Secondly, we have found that the scale from Weyl nodes' position in
the energy-momentum space can enter the anomaly equations. Finally,
we have fixed the ambiguity in the Nieh-Yan anomaly as well as other
cut-off dependence anomaly from the well-established momentum anomaly
in the $^{3}\text{He-A}$ phase. 

The rest of this paper is organized as follow. In Sec. \ref{sec:model},
we have introduced the effective model. In Sec. \ref{sec:anomalies_susy},
we have presented both the $U\left(1\right)$-gauge anomaly and the
diffeomorphism anomaly obtained from the supersymmetric quantum mechanics.
The non-conservation laws for both the $U\left(1\right)$-current
and the energy-momentum current are found, which are used to construct
the bulk-dislocation correspondence based on the Callan-Harvey mechanism.
In Sec. \ref{sec:spectral_flow_chiral}, the spectral flow calculations
are given and some subtleties in the effective model are clarified.
In Sec. \ref{sec:momentum_anomaly}, we have applied our results to
the momentum anomaly problem in the $^{3}\text{He-A}$ phase, where
we have fixed the ambiguities in the torsional anomalies due to the
cut-off dependence. In Sec. \ref{sec:conclusion}, the main results
of this paper are summarized. Finally, the supersymmetric quantum
mechanics calculations of anomalies and the conservation laws associated
with the covariant Lie derivative are given in details in Append.
\ref{sec:torsional_Lichnerowi} and Append. \ref{sec:susy_PI}, and
Append. \ref{sec:EM_conservation}. 

\section{Model \label{sec:model}}

We are interested in the Weyl Fermions in curved spacetime with torsion,
where the curved spacetime can emerge either from the order parameters
or the topological defects, including the dislocations and the disclinations
\citep{katanaev1992aop}. For concreteness, let us consider the following
model for the $^{3}\text{He-A}$ phase \citep{volovik2003oxford},
i.e., 
\begin{equation}
H=\frac{1}{2m}\left(\left|\boldsymbol{p}\right|^{2}-p_{F}^{2}\right)\tau^{3}+\tau^{1}\left(\boldsymbol{m}\cdot\boldsymbol{p}\right)+\tau^{2}\left(\boldsymbol{n}\cdot\boldsymbol{p}\right),\label{eq:He3-A}
\end{equation}
where $\tau^{i}$ with $i=1,\ 2,\ 3$ is the Pauli matrices and both
the vector $\boldsymbol{m}$ and $\boldsymbol{n}$ are from the order
parameters. Clearly, there exists gapless nodes located at $\boldsymbol{p}=\pm p_{F}\boldsymbol{l}$
in the dispersion relation, where $\boldsymbol{l}$ is defined as
$\boldsymbol{l}\equiv\frac{\boldsymbol{m}\times\boldsymbol{n}}{\left|\boldsymbol{m}\times\boldsymbol{n}\right|}$.
In the low-energy regime, the effective Hamiltonian can be written
as 
\[
H_{\pm}=\pm v_{F}\left(\boldsymbol{l}\cdot\boldsymbol{p}\right)\tau^{3}+\tau^{1}\left(\boldsymbol{m}\cdot\boldsymbol{p}\right)+\tau^{2}\left(\boldsymbol{n}\cdot\boldsymbol{p}\right)-v_{F}p_{F}\tau^{3},
\]
or up to an unitary rotation 
\begin{equation}
H=\gamma^{0}\gamma^{I}\left(\boldsymbol{e}_{I}\cdot\boldsymbol{p}-\gamma_{5}\lambda_{I}\right),\label{eq:Hamiltonian}
\end{equation}
where $I=1,$ 2, 3, $\gamma^{I}$ is the gamma matrix and the Fermi
velocity $v_{F}$ is $v_{F}\equiv p_{F}/m$. $\lambda_{I}=\delta_{I}^{3}v_{F}p_{F}$,
$\boldsymbol{e}_{I=3}=v_{F}\boldsymbol{l}$, $\boldsymbol{e}_{I=1}=\boldsymbol{m}$
and $\boldsymbol{e}_{I=2}=\boldsymbol{n}$. Similarly, for dislocations,
the Hamiltonian can be written in an analogue form \citep{sumiyoshi2016prl,huang2018prb}.
But now the vector $\boldsymbol{e}_{I}$ is from the lattice deformations
rather than the order parameters. Compared to the lattice without
dislocations, i.e., $e_{I}^{\mu}=\delta_{I}^{\mu}$, the position
of Weyl nodes is shifted. 

In addition to the torsion, we can further assume the existence of
curvature. Then, from Eq. (\ref{eq:Hamiltonian}), we can write down
the following action, 

\begin{equation}
S=\int d^{4}x\frac{1}{2}\left[\bar{\psi}\gamma^{a}\left(e_{a}^{\mu}iD_{\mu}-\lambda_{a}\gamma_{5}\right)\psi-\bar{\psi}\left(i\overleftarrow{D}_{\mu}e_{a}^{\mu}-\lambda_{a}\gamma_{5}\right)\gamma^{a}\psi\right],\label{eq:action}
\end{equation}
where $\mu=0,\ 1,\ 2,\ 3$ is the Einstein indices with coordinate
basis $\left\{ \partial_{\mu}\right\} $ and $a=0,\ 1,\ 2,\ 3$ is
the Lorentz indices with basis $\left\{ e_{a}^{\mu}\partial_{\mu}\right\} $.
$\gamma^{a}$ is the $4\times4$ gamma matrix, from which we can define
the generators of the Lorentz group, i.e., $\sigma^{ab}=\frac{1}{4}\left[\gamma^{a},\ \gamma^{b}\right]$.
$D_{\mu}$ and $\overleftarrow{D}_{\mu}$ are the Dirac operators,
i.e., 
\begin{equation}
D_{\mu}=\partial_{\mu}+\frac{1}{2}\omega_{ab\mu}\sigma^{ab}+iA_{\mu},
\end{equation}
and 
\begin{equation}
\overleftarrow{D}_{\mu}=\overleftarrow{\partial}_{\mu}-\frac{1}{2}\omega_{ab\mu}\sigma^{ab}-iA_{\mu},
\end{equation}
where $A_{\mu}$ is the $U\left(1\right)$ gauge fields and $\omega_{ab\mu}$
is the spin connection. $\omega_{ab}\equiv\omega_{ab\mu}dx^{\mu}$
can be regarded as the gauge field corresponding to the Lorentz group
with $\sigma^{ab}$ as the generators, whose curvature tensor is $\Omega_{ab}\equiv d\omega_{ab}+\left(\omega\wedge\omega\right)_{ab}$.
In addition to the curvature, we can also define the torsion, i.e.,
$T^{a}=de^{*a}+{\omega^{a}}_{b}\wedge e^{*b}$, where $e_{a}\equiv e_{a}^{\mu}\partial_{\mu}$
is the frame field, $e^{*a}\equiv e_{\mu}^{*a}dx^{\mu}$ is the co-frame
field and they satisfy $e_{\mu}^{*a}e_{b}^{\mu}=\delta_{b}^{a}$.
For later convenience, we also define $\tilde{T}^{a}\equiv de^{*a}$.
Then, in the absence of the spin connection, the torsion becomes ${T^{a}}_{\mu\nu}=\partial_{\mu}e_{\nu}^{*a}-\partial_{\nu}e_{\mu}^{*a}$,
so $T^{a}$ and $\tilde{T}^{a}$ coincide. 

Based on the action in Eq. (\ref{eq:action}), we are now ready to
derive the corresponding anomalies. Generally speaking, for a Weyl
Fermion under external gauge fields and gravity, there are
three kinds of anomalies, the gauge anomaly, the Einstein anomaly
(or the diffeomorphism anomaly) and the Lorentz anomalies, which relate
to the charge current, the energy-momentum current and the angular
momentum, respectively. We shall focus on the gauge anomaly as well
as the diffeomorphism anomaly, but the derivation of the angular momentum
is parallel. 

\section{Anomalies from supersymmetric quantum mechanics and Bulk-Dislocation
correspondence \label{sec:anomalies_susy}}

In this section, we shall derive both the $U\left(1\right)$ anomaly
and the diffeomorphism anomaly for Weyl Fermions. Then, these anomalies
are employed to construct the bulk-dislocation correspondence. 

Due to the descent relation, gauge anomalies in $d$-dimensional spacetime
are encoded in the chiral anomalies in $\left(d+2\right)$ dimensions
\citep{alvarez1985aop}. This can be straightforwardly appreciated
from the topological phase of matters. For concreteness and heuristic
purposes, let us focus on the electromagnetic responses. Consider
a $\left(3+1\right)$-dim time-reversal symmetry protected topological
insulator, whose effective action is known to be \citep{qi2008prb,witten2016rmp}
\begin{equation}
S_{\text{TI}}=-\frac{1}{8\pi}\int d^{4}x\epsilon^{\mu\nu\rho\sigma}\partial_{\mu}A_{\nu}\partial_{\rho}A_{\sigma}.\label{eq:TI}
\end{equation}
Since the effective model for the time-reversal symmetry protected
topological insulators is essentially the Dirac equation with negative
mass, the effective action in Eq. (\ref{eq:TI}) can be derived by
performing a chiral transformation to reverse the sign of the mass,
i.e., $\psi\rightarrow e^{i\frac{\pi}{2}\gamma_{5}}\psi$ and $\bar{\psi}\psi\rightarrow-\bar{\psi}\psi$.
That is, the chiral anomaly in four dimensions is encoded in the effective
action of the time-reversal invariant topological insulators. For
open manifolds with boundaries, Eq. (\ref{eq:TI}) is no longer time-reversal
invariant. The time-reversal symmetry can be preserved if the boundary
effective action is \citep{witten2016rmp}
\begin{equation}
S_{\text{PA}}^{\prime}=\frac{1}{8\pi}\int d^{3}x\epsilon^{\mu\nu\rho}A_{\mu}\partial_{\nu}A_{\rho}.\label{eq:parity_anomaly_massless}
\end{equation}
That is, in the presence of boundaries, both $S_{\text{PA}}^{\prime}$
and $S_{\text{TI}}$ are not time-reversal invariant, but their sum
is. This is guaranteed by the Atiyah-Patodi-Singer index theorem.
Especially, Eq. (\ref{eq:parity_anomaly_massless}) describes the
$\left(2+1\right)$-dim parity anomaly for the boundary Dirac Fermions.
Thus, we have seen the close relation between the $\left(d+1\right)$-dim
parity anomaly and the $\left(d+2\right)$-dim chiral anomaly. 

Now let us turn to the $\left(2+1\right)$-dim Dirac Fermions with
mass $m$, which is the effective model for Chern insulators. Analogue
to Eq. (\ref{eq:parity_anomaly_massless}), its effective action is 

\begin{equation}
S_{\text{PA}}=\frac{\text{sgn}\left(m\right)}{8\pi}\int d^{3}x\epsilon^{\mu\nu\rho}A_{\mu}\partial_{\nu}A_{\rho}.\label{eq:parity_anomaly}
\end{equation}
Note that for Chern insulators in the lattice, the prefactor should
be $\frac{1-\text{sgn}\left(m\right)}{2}$, but we shall focus on
the continuum model instead. As for the $U\left(1\right)$ gauge anomaly
for Weyl Fermions with chirality $s=\pm1$ in $d$ dimensions, it
can be understood from Eq. (\ref{eq:parity_anomaly}) by using the
Callan-Harvey mechanism \citep{callan1985npb}. For the $\left(2+1\right)$-dim
massive Dirac Fermions with a domain wall, it is well-known that there
exist $\left(1+1\right)$-dim chiral zero modes living in the domain
wall. Under external electromagnetic fields, the Hall current flows
from the bulk to the domain wall \citep{callan1985npb,stone1991aop},
i.e., 
\[
j^{\mu}=2\frac{s}{4\pi}\epsilon^{\mu\nu\rho}\partial_{\nu}A_{\rho},
\]
where $s=\text{sgn}\left(m\right)$ is the chirality of the Weyl Fermions
in the domain wall and $2$ is from the bulk Dirac Fermions in the
two sides of the domain wall. The bulk currents inflow leads to non-conservation
of currents in the domain wall, i.e., 
\begin{equation}
\partial_{\mu}j^{\mu}=\frac{s}{2\pi}\epsilon^{\mu\nu}\partial_{\mu}A_{\nu},
\end{equation}
which is the covariant $U\left(1\right)$ anomaly for a Weyl Fermion
with chirality $s$ \citep{alvarez1984npb,stone2012prb,parrikar2014prd}.
This completes the descent relation between the chiral anomaly, the
parity anomaly and the gauge anomaly. In addition to the gauge fields,
this construction can be easily generalized to include the gravitational
fields. Alternatively, one can construct the relation between the
$\left(d+2\right)$-dim chiral anomaly, the $\left(d+1\right)$-dim
parity anomaly and the $d$-dim gauge anomaly by considering the adiabatic
limit \citep{alvarez1984npb,parrikar2014prd}. 

The $\left(3+1\right)$-dim $U\left(1\right)$ anomaly and diffeomorphism
anomaly for Weyl Fermions can thus be derived from the chiral anomaly
in $\left(5+1\right)$-dim, or equivalently, the $\left(4+1\right)$-dim
parity-odd effective action. This can be done by employing the supersymmetric
quantum mechanics \citep{alvarez1983cmp,friedan1985pd,boer1996npb,bastianelli2006cambridge}.
Since the calculations are tedious and technical, we shall leave them
in Append. \ref{sec:torsional_Lichnerowi} and Append. \ref{sec:susy_PI}.
Because the physics is rather illustrating in the $\left(4+1\right)$-dim
parity-odd effective action (or, parity anomaly), we shall present
this effective action here, instead of the $\left(5+1\right)$-dim
Jacobian of the chiral transformation (see Eq. (\ref{eq:Jacobian_1})
in Append. \ref{sec:susy_PI}). The $\left(4+1\right)$-dim parity-odd
effective action is 
\begin{equation}
S_{\text{odd}}^{s\left(4+1\right)}=S_{\text{odd,\ }0}^{s}+\Delta S_{\text{odd}}^{s},\label{eq:effective_action_full}
\end{equation}
where $s=\pm1$ is the chirality of the boundary zero mode. $S_{\text{odd},\ 0}^{s}$
stands for the effective action independent of $\lambda_{a}^{s}$,
i.e., 

\begin{eqnarray}
sS_{\text{odd},\ 0}^{s} & = & -\frac{1}{8\pi^{2}\beta}\int A\wedge N-\frac{1}{24\pi^{2}}\int A\wedge F\wedge F\nonumber \\
 &  & -\frac{1}{192\pi^{2}}\int A\wedge\text{tr}\left(\widehat{\Omega}^{-}\wedge\widehat{\Omega}^{-}\right),\label{eq:effective_action_1}
\end{eqnarray}
where $N\equiv T^{a}\wedge T^{b}\eta_{ab}-e^{*a}\wedge e^{*b}\wedge\Omega_{ab}$
is the Nieh-Yan term, i.e., $\frac{1}{4!}N_{\mu\nu\rho\sigma}=\left(\frac{1}{2!}{T_{\ }^{a}}_{\mu\nu}\right)\left(\frac{1}{2!}{T_{\ }^{b}}_{\rho\sigma}\right)\eta_{ab}-\frac{1}{2!}e_{\rho}^{*a}e_{\sigma}^{*b}\Omega_{ab,\ \mu\nu}$.
$\widehat{\Omega}_{ab}^{-}$ is the curvature tensor associated with
$\widehat{\omega}_{ab\mu}^{-}\equiv\omega_{ab\mu}+{T^{\rho}}_{\mu\rho}+\frac{2}{3}H_{ab\mu}$
and $H\equiv e^{*a}\wedge T^{b}\eta_{ab}$, i.e., $\widehat{\Omega}_{ab}^{-}=d\widehat{\omega}_{ab}^{-}+\left(\widehat{\omega}^{-}\wedge\widehat{\omega}^{-}\right)_{ab}$.
Note that $\beta^{-1/2}$ is the cut-off, but not the temperature.

$\Delta S_{\text{odd}}^{s}$ denotes the action containing $\lambda_{a}^{s}$,
i.e., 

\begin{eqnarray}
s\Delta S_{\text{odd}}^{s} & = & -\frac{1}{8\pi^{2}\beta}\int\lambda_{a}^{s}e^{*a}\wedge N\nonumber \\
 &  & -\frac{1}{24\pi^{2}}\int\left(\lambda_{a_{1}}^{s}e^{*a_{1}}\right)\wedge\left(\lambda_{a_{2}}^{s}\tilde{T}^{a_{2}}\right)\wedge\left(\lambda_{a_{3}}^{s}\tilde{T}^{a_{3}}\right)\nonumber \\
 &  & -\frac{1}{8\pi^{2}}\int\left(\lambda_{a}^{s}e^{*a}\right)\wedge F\wedge F\nonumber \\
 &  & -\frac{1}{8\pi^{2}}\int\left(\lambda_{a}^{s}e^{*a}\right)\wedge\left(\lambda_{a}^{s}\tilde{T}^{a}\right)\wedge F\nonumber \\
 &  & -\frac{1}{192\pi^{2}}\int\lambda_{a_{1}}^{s}e^{*a_{1}}\wedge\text{tr}\widehat{\Omega}^{-}\wedge\widehat{\Omega}^{-},\label{eq:effective_action_2}
\end{eqnarray}
where $\tilde{T}^{a}\equiv de^{*a}$ is different from the torsion.
$\Delta S_{\text{odd}}^{s}$ is obtained here for the first time.
Compared to the parity-odd effective action for the Dirac Fermions
in five dimensions, the one in Eq. (\ref{eq:effective_action_full})
is different by a factor $2$, which is because we have compactified
the momentum space. In addition, in the derivation of the effective
action in Eq. (\ref{eq:effective_action_full}), we have neglected
terms of higher-order derivative in the chiral anomaly. The full expression
for the chiral anomaly is given in Eq. (\ref{eq:jacobian_chiralanomaly})
in Append. \ref{sec:susy_PI}. 

The covariant anomalies of the four dimensional boundary can be derived
by using the Callan-Harvey mechanism. Namely, the non-conservation
of the boundary charge currents and energy-momentum currents stem
from the bulk current inflow. Because $j_{s}^{\mu}\equiv-\frac{\delta}{\delta A_{\mu}}S_{\text{odd}}^{s}$
and $\tau_{s,\ a}^{\mu}\equiv-\frac{1}{\sqrt{\left|g\right|}}\frac{\delta}{\delta e_{\mu}^{*a}}S_{\text{odd}}^{s}$,
by performing variation upon $A_{\mu}$ and $e_{\mu}^{*a}$, respectively,
one can obtain the bulk currents and thus the boundary anomalies.
The bulk $U\left(1\right)$-current leads to the $\left(3+1\right)$-dim
covariant $U\left(1\right)$ anomaly on the boundary, i.e., 
\begin{equation}
\partial_{\mu}\left(\sqrt{\left|g\right|}j_{s}^{\mu}\right)=\mathcal{P}_{0,\ s}+\Delta\mathcal{P}_{s},\label{eq:u(1)_anomaly}
\end{equation}
where $\mathcal{P}_{0,\ s}$ contains terms from $S_{\text{odd},\ 0}^{s}$,
i.e., 
\begin{eqnarray}
 &  & \mathcal{P}_{0,\ s}\nonumber \\
 & = & \frac{s}{32\pi^{2}}\epsilon^{\mu\nu\rho\sigma}F_{\mu\nu}F_{\rho\sigma}\nonumber \\
 &  & +\frac{s}{32\pi^{2}\beta}\epsilon^{\mu\nu\rho\sigma}\left(\eta_{ab}{T^{a}}_{\mu\nu}{T^{b}}_{\rho\sigma}-2e_{\mu}^{*a}e_{\nu}^{*b}\Omega_{ab,\ \rho\sigma}\right)\nonumber \\
 &  & +\frac{s}{768\pi^{2}}\epsilon^{\mu\nu\rho\sigma}\widehat{\Omega}_{ab,\ \mu\nu}^{-}{{\widehat{\Omega}_{\ }^{-}}{}^{b}}{}_{a,\ \rho\sigma},\label{eq:j_s0}
\end{eqnarray}

and $\Delta\mathcal{P}_{s}$ from $\Delta S_{\text{odd}}^{s}$, i.e.,
\begin{equation}
\Delta\mathcal{P}_{s}=s\frac{\epsilon^{\mu\nu\rho\sigma}}{16\pi^{2}}\lambda_{a}^{s}{\tilde{T}^{a}}{}_{\mu\nu}F_{\rho\sigma}+s\frac{\epsilon^{\mu\nu\rho\sigma}}{32\pi^{2}}\lambda_{a}^{s}\lambda_{b}^{s}{\tilde{T}_{\ }^{a}}{}_{\mu\nu}{\tilde{T}_{\ }^{b}}{}_{\rho\sigma}.\label{eq:Delta_j_s}
\end{equation}

Terms in the Eq. (\ref{eq:j_s0}) match with those in Ref. \citep{parrikar2014prd}.
The term in the second line is the celebrated Adler-Bell-Jackiw anomaly
\citep{adler1969pr,bell1969nca}. This current non-conservation is
due to the charge pumping through the Weyl nodes from the Dirac sea
\citep{nielsen1983plb}, so it is captured by the low-energy effective
model and it is not sensitive to the ultra-violet physics. The term
in the third line in Eq. (\ref{eq:j_s0}) is known as the Nieh-Yan
anomaly \citep{nieh1982aop,nieh1982jmp,yajima1985ptp,chandia1997prd,obukhov1997fop,peeters1999jhep,Chandia1998prd}.
From the point of view of spectral flow, this anomaly is from the
particle inflow through the ultra-violet cut-off, which is dramatically
different from the Adler-Bell-Jackiw anomaly. The last term is the
mixed axial--gravitational anomaly \citep{alvarez1984npb}. However,
the curvature ${\widehat{\Omega}^{-}}_{ab,\ \mu\nu}$ is different
from $\Omega_{ab,\ \mu\nu}$ by the torsion tensor. 

Eq. (\ref{eq:Delta_j_s}) matches with the results from the semiclassical
approach \citep{huang2018prb}. Note that it is ${\tilde{T}^{a}}{}_{\mu\nu}$
appears here, but not the torsion ${T^{a}}_{\mu\nu}$. In the zero
spin connection limit, these two tensors coincide, but the physical
origins of the $\lambda_{a}^{s}\lambda_{b}^{s}{\tilde{T}_{\ }^{a}}{}_{\mu\nu}{\tilde{T}_{\ }^{b}}{}_{\rho\sigma}$
term and the Nieh-Yan term are very different. Terms in Eq. (\ref{eq:Delta_j_s})
are from the charge pumping through the Weyl nodes, while the Nieh-Yan
term is from the pumping through the ultra-violet cut-off. We shall
further explore this in Sec. \ref{sec:spectral_flow_chiral}. Finally,
in reality, Weyl nodes with different chiralities can locate at the
opposite position in the momentum space, i.e., $s\lambda_{a}$ for
$s$-Weyl Fermions. Then, the first term in Eq. (\ref{eq:Delta_j_s})
seems to imply charge non-conservation, i.e., 
\begin{equation}
\partial_{\mu}\left(\sqrt{\left|g\right|}j^{\mu}\right)=\frac{\epsilon^{\mu\nu\rho\sigma}}{8\pi^{2}}\lambda_{a}{\tilde{T}_{\ }^{a}}{}_{\mu\nu}F_{\rho\sigma},\label{eq:charge_non_conservation}
\end{equation}
where $j^{\mu}\equiv\sum_{s}j_{s}^{\mu}$. This non-conservation can
be understood from the Callan-Harvey mechanism. For concreteness,
let us consider the Weyl semimetals with a dislocation located at
$\left(x=0,\ y=0\right)$ along the $z$-axis, i.e., $\int{\tilde{T}_{\ }^{a}}{}_{xy}dxdy=b^{a}$
with $b^{a}$ the Burgers vector. We can thus approximate ${\tilde{T}_{\ }^{a}}{}_{xy}$
by the Dirac-Delta function, i.e., ${\tilde{T}_{\ }^{a}}{}_{xy}=b^{a}\delta\left(x\right)\delta\left(y\right)$.
Eq. (\ref{eq:charge_non_conservation}) becomes $\partial_{\mu}\left(\sqrt{\left|g\right|}j^{\mu}\right)=\frac{1}{2\pi^{2}}\lambda_{a}b^{a}\delta\left(x\right)\delta\left(y\right)E_{z}$
and $E_{z}$ is the electric field along the $z$-direction, which
must be canceled by the current non-conservation in the dislocations.
Hence, this anomaly equation means that there are chiral zero modes
trapped in the dislocations, whose chirality depends on the product
of the Burgers vector and $\lambda_{a}$. To be more specific, under
external electric fields, the chiral zero modes in the dislocations
lead to charge pumped up from the Dirac sea, which originates from
the bulk.

In addition, Eq. (\ref{eq:charge_non_conservation}) implies that
the corresponding effective action is 
\begin{equation}
S_{\text{eff}}^{\left(3+1\right)}=\int d^{4}x\frac{\epsilon^{\mu\nu\rho\sigma}}{4\pi^{2}}\lambda_{a}e_{\mu}^{*a}A_{\nu}\partial_{\rho}A_{\sigma},\label{eq:s_eff}
\end{equation}
which is from $\langle j^{\mu}\rangle\equiv-\frac{\delta}{\delta A_{\mu}}S_{\text{eff}}+\dots$
and $\sqrt{\left|g\right|}\langle j^{\mu}\rangle=\frac{\epsilon^{\mu\nu\rho\sigma}}{2\pi^{2}}\lambda_{a}e_{\nu}^{*a}\partial_{\rho}A_{\sigma}+\dots$
with ``$\dots$'' standing for terms satisfying the current conservation
law. Under gauge transformation $A_{\mu}^{\prime}=A_{\mu}+\partial_{\mu}\theta$,
there is $\delta S_{\text{eff}}^{\left(3+1\right)}=\frac{1}{4\pi^{2}}\int d^{4}x\epsilon^{\mu\nu\rho\sigma}\theta\lambda_{a}\partial_{\mu}e_{\nu}^{*a}\partial_{\rho}A_{\sigma}$,
so it is not gauge invariant. By requiring the gauge invariance, the
effective action for physics within the dislocations must cancel the
gauge non-invariance terms in Eq. (\ref{eq:s_eff}), i.e., $\delta S_{\text{eff}}^{\left(1+1\right)}=-\left(\frac{\lambda_{a}b^{a}}{\pi}\right)\left(\frac{1}{2\pi}\int\theta\epsilon^{\rho\sigma}\partial_{\rho}A_{\sigma}\right)$.
Notice that $\left(\frac{1}{2\pi}\int\theta\epsilon^{\rho\sigma}\partial_{\rho}A_{\sigma}\right)$
is the $U\left(1\right)$-gauge anomaly for Weyl Fermions, so there
must exist chiral zero modes trapped in the dislocations and their
chirality is determined by the sign of $-\frac{\lambda_{a}b^{a}}{\pi}$.
Suppose that $\lambda_{a}$ equals to the reciprocal vector, then
we have effectively obtained the weak topological insulators with
the stacking direction along $\lambda_{a}$. So Eq. (\ref{eq:s_eff})
can be used to describe the weak topological insulators and it matches
with that in Ref. \citep{nissinen2019prr}. 

As for the energy-momentum tensor, it can be obtained by variating
upon the vielbein, where the spin connection is kept fixed. That is,
we have taken the vielbein $e_{\mu}^{*a}$ and the spin connection
$\omega_{ab\mu}$ as independent variables. The reason is that for
the $^{3}\text{He-A}$ phase, the chiral superconductors/superfluids
and the Weyl semimetals, we wish to set the spin connection $\omega_{ab\mu}$
to zero regardless of the value of the vielbeins. By variating upon
the vielbein, one can obtain

\begin{eqnarray}
 &  & \left(\nabla_{\mu}+{T^{\rho}}_{\mu\rho}\right)\tau_{s,\ a}^{\mu}-\left(\tau_{s,\ b}^{\nu}e_{a}^{\mu}{T^{b}}_{\mu\nu}-S_{s}^{\nu cd}e_{a}^{\mu}\Omega_{cd,\ \mu\nu}-e_{a}^{\rho}F_{\rho\sigma}j^{\sigma}\right)\nonumber \\
 & = & \mathcal{Q}_{s,\ a}^{0}+\Delta\mathcal{Q}_{s,\ a},\label{eq:diffeomorphism_anomaly}
\end{eqnarray}
where $\tau_{s,\ a}^{\mu}\equiv-\frac{1}{\sqrt{\left|g\right|}}\frac{\delta S_{s}}{\delta e_{\mu}^{*a}}$
and $S_{s}^{\mu cd}\equiv\frac{1}{\sqrt{\left|g\right|}}\frac{\delta S_{s}}{\delta\omega_{cd\mu}}$
are the energy-momentum tensor and the spin current for the $s$-Weyl
Fermions, respectively. Terms on the first line of Eq. (\ref{eq:diffeomorphism_anomaly})
are from the classical equations of motion, or the Noether current
associated with the covariant Lie derivative (for details, please
refer to Sec. \ref{sec:EM_conservation}), i.e., 
\[
\delta_{\xi}^{C}e_{\nu}^{*a}=\xi^{\mu}{T^{a}}_{\mu\nu}+\nabla_{\nu}\xi^{a},
\]
\[
\delta_{\xi}^{C}\omega_{ab\nu}=\Omega_{ab,\ \mu\nu}\xi^{\mu},
\]
and 
\[
\delta_{\xi}^{C}A_{\nu}=\xi^{\mu}F_{\mu\nu}.
\]
This transformation is obtained by performing the Lie derivative associated
with the vector $\xi$, then a rotation $\left(i_{\xi}\omega\right)_{ab}$
and finally a $U\left(1\right)$ gauge transformation, $\exp\left(i\xi^{\mu}A_{\mu}\right)$.
$i_{\xi}$ is the interior derivative for the vector field $\xi$.
 In addition, Eq. (\ref{eq:diffeomorphism_anomaly}) can be recast
to a more illustrating form by introducing the Killing vectors. Because
the vielbeins and the spin connections are now independent variables,
we can define the generalized Killing vectors as $\delta_{K}^{C}e_{\nu}^{*a}=0$
and $\delta_{K}^{C}\omega_{ab\nu}=0$. Clearly, for the metric, the
first condition implies $\mathcal{L}_{K}g_{\mu\nu}=0$, which coincides
with the one used in general relativity. In addition, we should keep
the gauge field invariant under the covariant Lie derivative as well,
i.e., $\delta_{K}^{C}A_{\mu}=0$. Finally, one can recast Eq. (\ref{eq:diffeomorphism_anomaly})
as 
\begin{equation}
\frac{1}{\sqrt{\left|g\right|}}\nabla_{\mu}\left(\sqrt{\left|g\right|}\tau_{s,\ a}^{\mu}K^{a}\right)=\left(\mathcal{Q}_{s,\ a}^{0}+\Delta\mathcal{Q}_{s,\ a}\right)K^{a},\label{eq:diffeomorphism_anomaly_current}
\end{equation}
which becomes $\frac{1}{\sqrt{\left|g\right|}}\nabla_{\mu}\left(\sqrt{\left|g\right|}\tau_{s,\ a}^{\mu}K^{a}\right)=0$
at the classical limit and this is the energy-momentum current conservation. 

$\mathcal{Q}_{s,\ a}^{0}$ is from $S_{\text{odd},\ 0}^{s}$, i.e., 

\begin{eqnarray}
 &  & \mathcal{Q}_{s,\ a}^{0}\nonumber \\
 & = & \frac{s}{16\pi^{2}\beta}\frac{\epsilon^{\mu\nu\rho\sigma}}{\sqrt{\left|g\right|}}F_{\mu\nu}{T^{c}}_{\rho\sigma}\eta_{ac}\nonumber \\
 &  & -\frac{s}{144\pi^{2}}\frac{\epsilon^{\mu\nu\rho\sigma}}{\sqrt{\left|g\right|}}\eta_{ab}\left[\frac{1}{2}{T^{b}}_{\mu\nu}{\mathcal{T}_{2}^{\prime}}_{\rho\sigma}+\frac{1}{3!}\left(d\mathcal{T}_{2}^{\prime}\right)_{\mu\nu\rho}e_{\sigma}^{*b}\right],\label{eq:energy_momentum_s0}
\end{eqnarray}
where $\mathcal{T}_{2}^{\prime}\equiv i_{e_{a}}i_{e_{b}}\left(dA\wedge{\widehat{\Omega}^{-}}{}^{ba}\right)$
and $i_{e^{a}}$ is the interior derivative associated with the vector
$e_{a}$. 

$\Delta\mathcal{Q}_{s}$ is from $\Delta S_{\text{odd}}^{s}$, i.e., 

\begin{eqnarray}
 &  & \Delta\mathcal{Q}_{s,\ a}\nonumber \\
 & = & \lambda_{a}^{s}\left(\frac{1}{\sqrt{\left|g\right|}}\partial_{\mu}\sqrt{\left|g\right|}j_{s}^{\mu}\right)+\frac{s}{16\pi^{2}\beta}\frac{\epsilon^{\mu\nu\rho\sigma}}{\sqrt{\left|g\right|}}\eta_{ac}\left(\lambda_{b}^{s}{\tilde{T}_{\ }^{b}}{}_{\mu\nu}\right){T^{c}}_{\rho\sigma}\nonumber \\
 &  & -\frac{s}{144\pi^{2}}\frac{\epsilon^{\mu\nu\rho\sigma\alpha}}{\sqrt{\left|g\right|}}\eta_{ab}\left({T^{b}}_{\mu\nu}{\mathcal{T}_{2}}_{\rho\sigma}+\left(d\mathcal{T}_{2}\right)_{\mu\nu\rho}e_{\sigma}^{*b}\right),\label{eq:energy_momentum_delta}
\end{eqnarray}
where $\mathcal{T}_{2}\equiv i_{e_{a}}i_{e_{b}}\left(\lambda_{a_{1}}^{s}de^{*a_{1}}\wedge{\widehat{\Omega}^{-}}{}^{ba}\right)$. 

Eq. (\ref{eq:energy_momentum_s0}) matches with the results in Ref.
\citep{parrikar2014prd}. Terms in the second line can be understood
from the Landau levels \citep{huang2019arxiv_ham}. Specifically,
under external magnetic fields, the dimensions are effectively reduced
from $\left(3+1\right)$ to $\left(1+1\right)$. The torsional electric
fields ${T_{\ }^{a}}_{0i}$ would change the slope of the lowest Landau
levels. Consequently, there are particles squeezed out from the energy
cut-off, which leads to the energy-momentum anomaly. 

Terms in Eq. (\ref{eq:energy_momentum_delta}) are obtained here for
the first time. The physical meaning of the first term in Eq. (\ref{eq:energy_momentum_delta})
is straightforward. For the $s$-Weyl node locates at $\lambda_{a}^{s}$,
the averaged momentum of the excitations around this Weyl node is
$\lambda_{a}^{s}$. This provides an additional charge for these excitations,
the momentum $\lambda_{a}^{s}$. Consequently, the current non-conservation
in Eq. (\ref{eq:u(1)_anomaly}) leads to the energy-momentum non-conservation,
i.e., $\lambda_{a}^{s}\partial_{\mu}j_{s}^{\mu}$. 

Similar to the $U\left(1\right)$-current non-conservation, the last
term in the second line in Eq. (\ref{eq:energy_momentum_delta}) can
lead to energy-momentum non-conservation as well. This can be understood
from the Callan-Harvey mechansim. Namely, in the presence of dislocations
along the $z$-axis, ${\tilde{T}_{\ }^{a}}{}_{xy}$ is non-zero and
it can be approximated by ${\tilde{T}_{\ }^{a}}{}_{xy}=b^{a}\delta\left(x\right)\delta\left(y\right)$.
Consequently, the energy-momentum non-conservation becomes $\nabla_{\mu}\left(\tau_{a}^{\mu}K^{a}\right)=\left(\frac{\lambda_{d}b^{d}}{\pi}\right)\left(\frac{1}{4\pi^{2}\beta}\eta_{ac}\epsilon^{\rho\sigma}{T^{c}}_{\rho\sigma}K^{a}\right)+\dots$.
The energy-momentum non-conservation for the $\left(1+1\right)$-dim
Weyl Fermions is known to be $\nabla_{\mu}\left(\tau_{s,\ a}^{\mu}K^{a}\right)=\frac{s}{4\pi\beta}\epsilon^{\mu\nu}\eta_{ac}{T^{c}}_{\mu\nu}K^{a}$
\citep{hughes2013prd,huang2019arxiv_ham}. So this term originates
from the chiral zero modes trapped in the dislocations with chirality
decided by the sign of $-\frac{\lambda_{a}b^{a}}{\pi}$, which is
consistent with the analysis above. Especially, this anomaly term
involves only the torsion, but not the electromagnetic fields, so
it can be applied to the chiral superconductors as well. Because of
the particle-hole symmetry, there is an extra coefficient $\frac{1}{2}$
in the anomaly equations. In addition, the trapped chiral zero mode
is Majorana Fermion. 

\section{Spectral flow and chiral anomaly \label{sec:spectral_flow_chiral}}

In last section, we have obtained both the $U\left(1\right)$ anomaly
and the diffeomorphism anomaly from the supersymmetric quantum mechanic
approach. Interestingly, we have found that these two anomalies relate
to each other closely, i.e., $\nabla_{\mu}\tau_{s,\ a}^{\mu}=\lambda_{a}^{s}\left(\frac{1}{\sqrt{\left|g\right|}}\partial_{\mu}\sqrt{\left|g\right|}j_{s}^{\mu}\right)+\dots.$
In this section, we shall study the chiral anomaly from the view of
spectral flow. 

If both the spin connection and the electromagnetic fields are turned
to zero, then the chiral anomaly can be written as 

\begin{eqnarray}
 &  & \partial_{\mu}\left(\sqrt{\left|g\right|}j^{5\mu}\right)\nonumber \\
 & = & \frac{1}{16\pi^{2}\beta}\epsilon^{\mu\nu\rho\sigma}\eta_{ab}{T_{\ }^{a}}{}_{\mu\nu}{T_{\ }^{b}}{}_{\rho\sigma}+\frac{\epsilon^{\mu\nu\rho\sigma}}{16\pi^{2}}\lambda_{a}^{s}\lambda_{b}^{s}{\tilde{T}_{\ }^{a}}{}_{\mu\nu}{\tilde{T}_{\ }^{b}}{}_{\rho\sigma},\label{eq:anomaly_u1}
\end{eqnarray}
where we have only kept the torsional terms with lowest order of derivative.
Note that $\beta$ here is the cut-off, i.e., $\beta=\Lambda^{-2}$
rather than the temperature. Since the spin connection is zero, there
is ${T^{a}}_{\mu\nu}={\tilde{T}_{\ }^{a}}{}_{\mu\nu}$. Up to the
prefactors, terms in (\ref{eq:anomaly_u1}) look superficially the
same. However, as we shall show, they have totally different physical
origins. 

Let us first focus on the Nieh-Yan anomaly in Eq. (\ref{eq:anomaly_u1}).
The derivation of the zero-temperature Nieh-Yan anomaly in the absence
of vector $\lambda_{a}$ from the view of spectral flow has been provided
in Ref. \citep{parrikar2014prd,huang2019arxiv_thermal}. For consistency,
we shall review the calculations here. We first decompose the vielbein
as $e_{\mu}^{*a}=\delta_{\mu}^{a}+w_{\mu}^{a}$, where $w_{\mu}^{a}$
is assumed to be small. Then, we apply the torsional magnetic fields
along the $z$-direction, i.e., $w_{\mu}^{a}=\frac{1}{2}\delta_{3}^{a}\tilde{T}_{B}\left(0,\thinspace-y,\thinspace x,\thinspace0\right),\text{\ensuremath{\tilde{T}_{B}>0}}$,
where the translational symmetry along the $z$-direction is intact.
The Hamiltonian is thus given as 

\begin{equation}
H_{s}=s\left[\left(p_{x}+\frac{1}{2}\tilde{T}_{B}yp_{z}\right)\sigma^{1}+\left(p_{y}-\frac{1}{2}\tilde{T}_{B}xp_{z}\right)\sigma^{2}+p_{z}\sigma^{3}\right],
\end{equation}
whose dispersion relation is 
\[
\mathcal{E}_{s}^{n}=\begin{cases}
s\left|p_{z}\right| & n=0\\
\pm\sqrt{p_{z}^{2}+2\left|n\tilde{T}_{B}p_{z}\right|} & \left|n\right|>1
\end{cases}.
\]
Now we further turn on the torsional electric fields, i.e., $\tilde{T}_{E}^{*3}=\partial_{0}e_{z}^{*3}-\partial_{z}e_{0}^{*3}$.
For simplicity, we set $e_{0}^{*3}=0$, $e_{z}^{*3}=1+\Phi$ and $\Phi\ll1$.
The dispersion relation now becomes 
\begin{equation}
\mathcal{E}_{s}=\begin{cases}
\begin{array}{cc}
s\left(1-\Phi\right)\left|p_{z}\right| & n=0\\
\pm\sqrt{\left|\left(1-\Phi\right)p_{z}\right|^{2}+2\left|n\tilde{T}_{B}^{3}p_{z}\right|} & \left|n\right|\geq1
\end{array} & .\end{cases}
\end{equation}
 This shows that the torsional electric fields affect the lowest torsional
Landau levels by modifying its slope (or velocity), i.e., $\left|p_{z}\right|\rightarrow\left(1-\Phi\right)\left|p_{z}\right|$.
If we impose an energy cut-off, then by tuning the effective velocity,
some states are squeezed out from the infra-red regime to the ultra-violet
regime, which can lead to the Nieh-Yan anomaly. 

The axial density $j^{5\mu}|_{\mu=0}$ can be written as 
\begin{equation}
j^{50}=\sum_{n}\sum_{s}s\int_{-\infty}^{+\infty}\frac{dp_{z}}{2\pi}n_{F}\left(\mathcal{E}_{s}^{n}\right)\left(\frac{\left|p_{z}\right|\tilde{T}_{B}^{3}}{2\pi}\right),
\end{equation}
where $n_{F}\left(\mathcal{E}_{s}^{n}\right)=\frac{1}{\exp\left(\frac{1}{T}\mathcal{E}_{s}^{n}\right)+1}$
is the Fermi-Dirac distribution function, $T$ is the temperature
and $\frac{\left|p_{z}\right|\tilde{T}_{B}^{3}}{2\pi}$ is the level
degeneracy. For $\left|n\right|\geq1$, $\mathcal{E}_{R}^{n}=\mathcal{E}_{L}^{n}$,
there is $\sum_{\left|n\right|>1}\sum_{s}sn_{F}\left(\mathcal{E}_{s}^{n}\right)\left|p_{z}\right|=0$,
so their contributions to $j^{50}$ are canceled exactly. Now at the
zero-temperature limit, the chiral density becomes 

\begin{eqnarray*}
 &  & j^{50}/\left[\tilde{T}_{B}^{3}/\left(2\pi\right)\right]\\
 & = & \int_{-\infty}^{+\infty}\frac{dp_{z}}{2\pi}\left|p_{z}\right|\\
 &  & \times\left\{ \frac{1}{\exp\left[\frac{1}{T}\left(1-\Phi\right)\left|p_{z}\right|\right]+1}-\frac{1}{\exp\left[-\frac{1}{T}\left(1-\Phi\right)\left|p_{z}\right|\right]+1}\right\} \\
 & = & \frac{1}{\left(1-\Phi\right)^{2}}2\int_{-\Lambda}^{+\Lambda}\frac{d\epsilon}{2\pi}\epsilon\frac{1}{\exp\left(\frac{1}{T}\epsilon\right)+1}\\
 & = & -\frac{1}{2\pi}\Lambda^{2}-\frac{1}{\pi}\Lambda^{2}\Phi+\mathcal{O}\left(\Phi^{2}\right),
\end{eqnarray*}
where $\epsilon\equiv\left(1-\Phi\right)\left|p_{z}\right|$ and $\Lambda$
is the energy cut-off. The Nieh-Yan anomaly can be easily obtained
by recast $\partial_{t}j^{50}$ to a covariant form. Hence, from the
view of the spectral flow, the Nieh-Yan anomaly can be understood
as follow. The slope of the lowest torsional Landau levels is changed
by the torsional electric fields, i.e., $1\rightarrow\frac{1}{1-\Phi}$.
Because of the energy cut-off, some particles are squeezed out from
low-energy regime, when we adiabatically tune the slope of the lowest
torsional Landau levels. That is, in the Nieh-Yan anomaly, the extra
particles come from the ultra-violet region. By contrast, for the
Adler-Bell-Jackiw anomaly, the extra particles are from the infra-red
regime. Namely, under external electric fields, particles are pumped
up through the lowest Landau levels through the Weyl nodes. 

\begin{figure}
\includegraphics[scale=0.5]{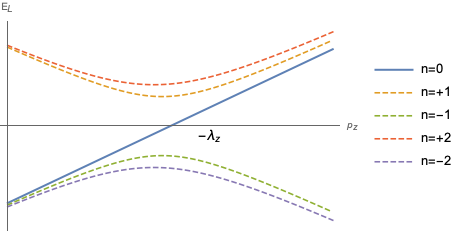}

\includegraphics[scale=0.5]{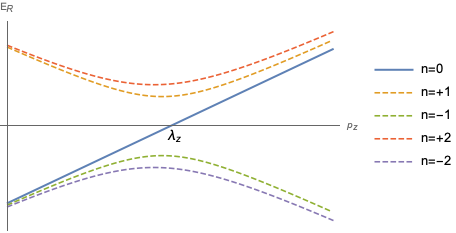}

\caption{Torsional Landau levels for the model in Eq. (\ref{eq:Hamiltonian_He3}). Top
panel: the torsional Landau levels for the left-handed Weyl Fermions
in the region $\lambda_{z}-\Lambda<p_{z}<\lambda_{z}+\Lambda$ in
the limit $\lambda_{z}/\Lambda\gg1$ where $\Lambda$ is the region
where the linear model is effective. Bottom panel: the torsional Landau
levels for the right-handed Weyl Fermions in the region $-\lambda_{z}-\Lambda<p_{z}<-\lambda_{z}+\Lambda$
in the limit $\lambda_{z}/\Lambda\gg1$. \label{fig:TLL_MA}}
\end{figure}

\begin{figure}
\includegraphics[scale=0.5]{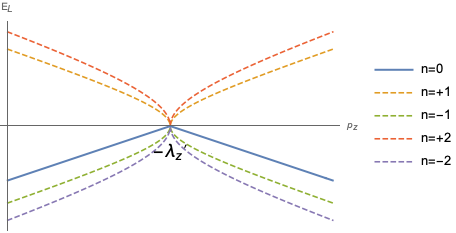}

\includegraphics[scale=0.5]{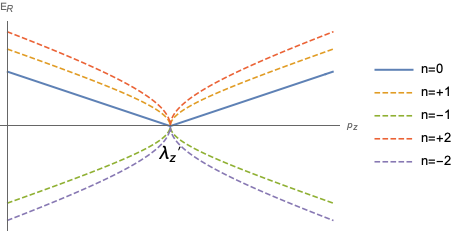}

\caption{Torsional Landau levels for the model in Eq. (\ref{eq:Hamiltonian_thermal}). Top panel: the torsional
Landau levels for the left-handed Weyl Fermions. Bottom panel: the
torsional Landau levels for the right-handed Weyl Fermions. \label{fig:TLL_NY}}
\end{figure}

Now we can split the Weyl nodes. For example, in the absence of vielbeins,
the Hamiltonian is $H_{s}=s\left(\boldsymbol{p}-s\boldsymbol{\lambda}\right)\cdot\sigma$.
But when the vielbein is applied, the Hamiltonian becomes
\begin{equation}
H_{s}=s\boldsymbol{e}_{I}\cdot\boldsymbol{p}\sigma^{I}-\boldsymbol{\lambda}\cdot\sigma,\label{eq:Hamiltonian_He3}
\end{equation}
or 
\begin{equation}
H_{s}^{\prime}=s\boldsymbol{e}_{I}\cdot\left(\boldsymbol{p}-s\boldsymbol{\lambda}^{\prime}\right)\sigma^{I},\label{eq:Hamiltonian_thermal}
\end{equation}
where both $\boldsymbol{\lambda}$ and $\boldsymbol{\lambda}^{\prime}$
are constant and for later convenience, we use $\boldsymbol{\lambda}$
and $\boldsymbol{\lambda}^{\prime}$ to distinguish them.
Note that in Eq. (\ref{eq:Hamiltonian_He3}), the positions of Weyl
nodes are shifted by the vielbein, while the Weyl nodes in Eq. (\ref{eq:Hamiltonian_thermal})
are not and they locate at $s\boldsymbol{\lambda}^{\prime}$. The
one in Eq. (\ref{eq:Hamiltonian_He3}) has been used to describe the
$^{3}\text{He-A}$ phase and the chiral superconductors/superfluids.
Under dislocations, the coordinates are shifted from $x$ to $x^{\prime}$,
i.e., $x^{\prime}=x+u$. The momentum operator is thus changed, i.e.,
$\partial_{x}\rightarrow\frac{\partial x}{\partial x^{\prime}}\partial_{x}$
and we can use Eq. (\ref{eq:Hamiltonian_He3}) as the effective model
for the dislocations. As for Eq. (\ref{eq:Hamiltonian_thermal}),
the vielbein couples with the relative momentum ($\boldsymbol{p}-s\boldsymbol{\lambda}^{\prime}$)
rather than the total momentum. Similarly, the coupling charge of
$e_{\mu}^{*a=0}$ should be $\left(p_{0}-s\mu\right)$. Hence, based
on Luttinger's method \citep{luttinger1964pr}, Eq. (\ref{eq:Hamiltonian_thermal})
is applicable for the thermal transport. However, in this paper, we
shall restrict ourselves to the zero-temperature phenomena. 

The torsional Landau levels for Eq. (\ref{eq:Hamiltonian_He3}) are
given as \citep{balatskii1986zetf,huang2018prb}

\begin{equation}
\mathcal{E}_{s}^{n}=\begin{cases}
\begin{cases}
s\left(p_{z}-s\lambda_{z}\right) & p_{z}>0\\
-s\left(p_{z}-s\lambda_{z}\right) & p_{z}<0
\end{cases} & n=0\\
\pm\sqrt{\left(p_{z}-\lambda_{z}\right)^{2}+2\left|n\tilde{T}_{B}p_{z}\right|} & \left|n\right|\geq1
\end{cases},\label{eq:dispersion_momentum_anomaly}
\end{equation}
where $\lambda_{a}\equiv\left(0,\ 0,\ 0,\ \lambda_{z}\right)$ and
$\lambda_{z}\geq0$. Since Eq. (\ref{eq:Hamiltonian_He3}) is only
applicable around $\boldsymbol{p}=s\boldsymbol{\lambda}$ and there
are energy gaps around $p_{z}=\pm\lambda_{z}$, we can assume $\left|\boldsymbol{\lambda}\right|$
is large enough, i.e., $\left|\boldsymbol{\lambda}\right|/\Lambda\gg1$
and $\Lambda$ is the region that the linear model in Eq. (\ref{eq:Hamiltonian_He3})
applies. The corresponding dispersion relation is shown in Fig. \ref{fig:TLL_MA}
and it reminds us of the magnetic Landau levels. Then, due to the
gap in the higher torsional Landau levels, we can calculate the Nieh-Yan
term in the chiral anomaly equation by focusing on the lowest torsional
Landau levels

\begin{eqnarray*}
 &  & j^{50}/\left[\tilde{T}_{B}^{3}/\left(2\pi\right)\right]\\
 & = & \lim_{T\rightarrow0}\int\frac{dp_{z}}{2\pi}\{p_{z}\frac{1}{\exp T^{-1}\left[\left(1-\Phi\right)p_{z}-\lambda_{z}\right]+1}\\
 &  & -\left(-p_{z}\right)\frac{1}{\exp T^{-1}\left[\left(1-\Phi\right)p_{z}+\lambda_{z}\right]+1}\}\\
 & = & -\frac{1}{2\pi}\Lambda^{2}-\frac{1}{\pi}\Lambda^{2}\Phi+\mathcal{O}\left(\Phi^{2}\right),
\end{eqnarray*}
where $\Lambda$ is the energy cut-off, i.e., $-\Lambda<\left(1-\Phi\right)p_{z}-\lambda_{z}<\Lambda$
and $-\Lambda<\left(1-\Phi\right)p_{z}+\lambda_{z}<\Lambda$ for the
right-handed and left-handed Weyl Fermions, respectively. The level
degeneracy is $\frac{\left|p_{z}\right|\tilde{T}_{B}}{2\pi}$, which
is $s\frac{p_{z}\tilde{T}_{B}}{2\pi}$ for the right-handed and left-handed
Weyl Fermions, respectively. Finally, by recasting $\partial_{t}j^{50}$
to a covariant form, one can obtain the Nieh-Yan anomaly. It is clear
that the Nieh-Yan anomaly is from the particles inflow from the energy
cut-off, rather than the Weyl nodes. In addition, the coefficient
is the energy density of the torsional Landau levels.

In addition to the Nieh-Yan anomaly above, there are particles pumped
up through the Weyl nodes. To appreciate this, we rewrite $\Phi$
as $\Phi=\tilde{T}_{E}\Delta t$.
The chiral density is

\begin{eqnarray*}
 &  & j^{50}/\left[\tilde{T}_{B}^{3}/\left(2\pi\right)\right]\\
 & = & \lim_{T\rightarrow0}\int\frac{dp_{z}}{2\pi}\{p_{z}\frac{1}{\exp T^{-1}\left[\left(1-\tilde{T}_{E}\Delta t\right)p_{z}-\lambda_{z}\right]+1}\\
 &  & -\left(-p_{z}\right)\frac{1}{\exp T^{-1}\left[\left(1-\tilde{T}_{E}\Delta t\right)p_{z}+\lambda_{z}\right]+1}\}\\
 & \simeq & \lim_{T\rightarrow0}\int\frac{dp_{z}}{2\pi}\left(-\tilde{T}_{E}\Delta tp_{z}^{2}\right)\{\partial_{\epsilon}\frac{1}{\exp T^{-1}\left(p_{z}-\lambda_{z}\right)+1}\\
 &  & +\partial_{\epsilon}\frac{1}{\exp T^{-1}\left(p_{z}+\lambda_{z}\right)+1}\}+\text{Const}\\
 & = & \lambda_{z}^{2}\frac{\tilde{T}_{E}}{\pi}\Delta t+\text{Const},
\end{eqnarray*}
where ``$\text{Const}$'' stands for the constant terms. In the
fourth-line, we have perform Taylor's expansion of the Fermi-Dirac
distribution function. By recast $\partial_{t}j^{50}$ to a covariant
form, one can obtain $\partial_{\mu}\sqrt{\left|g\right|}j^{5\mu}=\frac{\epsilon^{\mu\nu\rho\sigma}}{16\pi^{2}}\lambda_{a}\lambda_{b}\partial_{\mu}e_{\nu}^{*a}\partial_{\rho}e_{\sigma}^{*b}$.
Due to the Dirac-Delta function from the derivative of the Fermi-Dirac
distribution, this anomaly term is from particles pumped up through
the Weyl nodes, which is different from the Nieh-Yan anomaly. Alternatively,
one can derive this anomaly from the semiclassical approach \citep{huang2018prb},
where the coefficient of this anomaly equation is shown to be the
monopole charge in the momentum space. 

The Landau levels for Eq. (\ref{eq:Hamiltonian_thermal}) is shown
in Fig. (\ref{fig:TLL_NY}). Similar to the one with $\lambda_{a}=0$,
one can derive the Nieh-Yan anomaly from the lowest torsional Landau
levels. However, the $\frac{\epsilon^{\mu\nu\rho\sigma}}{16\pi^{2}}\lambda_{a}\lambda_{b}\partial_{\mu}e_{\nu}^{*a}\partial_{\rho}e_{\sigma}^{*b}$
term is zero in this case. This is because the coupling charge of
vielbein is $\left(p_{z}\mp\lambda_{z}\right)$ rather than $p_{z}$.
Due to the Dirac-Delta function ($\delta\left(p_{z}\mp\lambda_{z}\right)$)
from the Fermi-Dirac distribution, the coupling charge becomes zero. 

\section{Nieh-Yan anomaly in $^{3}\text{He-A}$ phase \label{sec:momentum_anomaly}}

In this section, we shall apply our results to the $^{3}\text{He-A}$
phase, where the momentum anomaly can be easily understood from our
anomaly equations. In addition, by comparing our results with the
well-established ones in the $^{3}\text{He-A}$, we can fix the ambiguities
in the anomaly equation from cut-off dependence, for example the Nieh-Yan
anomaly. 

In Ref. \citep{mermin1980prb,combescot1986prb,balatskii1986zetf,volovik1986jetp,stone1987aop},
based on different approaches, the following momentum anomaly is found
for the $^{3}\text{He-A}$ phase, 
\begin{equation}
\partial_{t}g_{j}+\partial_{i}{\pi^{i}}_{j}=\frac{3}{2}\rho l_{j}\left[\partial_{t}\boldsymbol{l}\cdot\left(\nabla\times\boldsymbol{l}\right)\right],\label{eq:momentum_anomaly}
\end{equation}
where $g_{i}$ is the mass current (or momentum density), ${\pi^{i}}_{j}$
is the momentum tensor and $\rho$ is the density of the superfluid,
i.e., $\rho=\frac{1}{6\pi^{2}}p_{F}^{3}$. In the model given in Eq.
(\ref{eq:He3-A}), there are $\lambda_{a}^{s}=\delta_{a}^{3}sv_{F}p_{F}$,
$\boldsymbol{e}_{a=1}=\boldsymbol{m}$, $\boldsymbol{e}_{a=2}=\boldsymbol{n}$
and $\boldsymbol{e}_{a=3}=v_{F}\boldsymbol{l}$. This means that terms
on the right-handed side of Eq. (\ref{eq:momentum_anomaly}) can be
recast as 
\[
\frac{1}{2}\frac{1}{2\pi^{2}}p_{F}^{3}l_{j}\left[\partial_{t}\boldsymbol{l}\cdot\left(\nabla\times\boldsymbol{l}\right)\right]=\frac{1}{2}\frac{\epsilon^{\mu\nu\rho\sigma}}{32\pi^{2}}\sum_{s=\pm}se_{j}^{*a}\lambda_{a}^{s}\left(\lambda_{c}^{s}{\tilde{T}^{c}}_{\mu\nu}\right)\left(\lambda_{d}^{s}{\tilde{T}^{d}}_{\rho\sigma}\right),
\]
so in our language, Eq. (\ref{eq:momentum_anomaly}) can be written
as 
\begin{eqnarray}
 &  & \frac{1}{\sqrt{\left|g\right|}}\nabla_{\mu}\left(\sqrt{\left|g\right|}{\tau^{\mu}}_{\nu}K^{\nu}\right)\nonumber \\
 & = & \frac{1}{2}\frac{\epsilon^{\mu\nu\rho\sigma}}{32\pi^{2}}\sum_{s=\pm}s\left(K^{\alpha}e_{\alpha}^{*a}\lambda_{a}^{s}\right)\left(\lambda_{c}^{s}{\tilde{T}^{c}}_{\mu\nu}\right)\left(\lambda_{d}^{s}{\tilde{T}^{d}}_{\rho\sigma}\right),\label{eq:momentum_anomaly_torsion}
\end{eqnarray}
where ${\tau^{\mu}}_{\nu}\equiv e_{\nu}^{*a}\tau_{a}^{\mu}$ and $K^{\nu}$
is the generalized Killing vector. 

Now we can compare the result obtained above with the momentum anomaly
in Eq. (\ref{eq:momentum_anomaly_torsion}). By setting the spin connection
to zero and considering the long-wavelength and low-energy limit,
the energy-momentum current in Eq. (\ref{eq:diffeomorphism_anomaly_current})
satisfies

\begin{eqnarray}
 &  & \frac{1}{\sqrt{\left|g\right|}}\nabla_{\mu}\left(\sqrt{\left|g\right|}\tau_{a}^{\mu}K^{a}\right)\nonumber \\
 & = & \frac{1}{2}\sum_{s}sK^{a}[\frac{1}{32\pi^{2}}\frac{\epsilon^{\mu\nu\rho\sigma}}{\sqrt{\left|g\right|}}\lambda_{a}^{s}\left(\lambda_{c}^{s}{\tilde{T}^{c}}_{\mu\nu}\right)\left(\lambda_{d}^{s}{\tilde{T}^{d}}_{\rho\sigma}\right)\nonumber \\
 &  & +\frac{1}{32\pi^{2}\beta}\frac{\epsilon^{\mu\nu\rho\sigma}}{\sqrt{\left|g\right|}}\lambda_{a}^{s}\left(\eta_{bc}{T^{b}}_{\mu\nu}{T^{c}}_{\rho\sigma}\right)\nonumber \\
 &  & +\frac{1}{16\pi^{2}\beta}\frac{\epsilon^{\mu\nu\rho\sigma}}{\sqrt{\left|g\right|}}\eta_{ac}\left(\lambda_{b}^{s}\tilde{T}_{\mu\nu}^{b}\right){T^{c}}_{\rho\sigma}],\label{eq:momentum_anomaly_susy}
\end{eqnarray}
where the external electromagnetic fields are set to zero due to the
particle-hole symmetry and ${\tilde{T}_{\ }^{a}}{}_{\mu\nu}\equiv\partial_{\mu}e_{\nu}^{*a}-\partial_{\nu}e_{\mu}^{*a}$
coincides with the torsion ${T^{a}}_{\mu\nu}$. The prefactor $\frac{1}{2}$
is because we are considering the Majorana Fermions instead of the
Dirac Fermions. The term in the second line is the momentum anomaly
shown in Eq. (\ref{eq:momentum_anomaly}) and it is known in the literatures
\citep{combescot1986prb,stone1985prl,volovik1985jetp_normal,balatskii1986zetf,volovik1986jetp,volovik1986jetp_gravity,stone1987aop,bevan1997nature}.
Some of these previous calculations for the momentum-current non-conservation
in the $^{3}\text{He}-A$ are not based on the low-energy effective
model around the gapless Weyl nodes, for example, Ref. \citep{cross1975jltp,mermin1980prb,combescot1986prb,stone1987aop}.
Especially, the term in the second line in Eq. (\ref{eq:momentum_anomaly_susy})
is also confirmed by experimental measurement \citep{bevan1997nature}.
Hence, one can conclude that the last two terms in Eq. (\ref{eq:momentum_anomaly_susy})
are negligible compared to the first term. This is because the energy
scale $\Lambda$ of the effective model in Eq. (\ref{eq:Hamiltonian})
is relatively small compared to $p_{F}$. To be more specific, let
us recast the paring order parameter as $\left(\boldsymbol{m}+i\boldsymbol{n}\right)=c_{\perp}\left(\hat{\boldsymbol{m}}+i\hat{\boldsymbol{n}}\right)$.
The scale $mc_{\perp}^{2}$ mark the border where the emergent Lorentz
symmetry breakdown \citep{volovik2003oxford}. For example, in the
limit $m\rightarrow0$, the $\frac{1}{2m}\left(\left|\boldsymbol{p}\right|^{2}-p_{F}^{2}\right)$
term in Eq. (\ref{eq:He3-A}) dominates and the system shows non-relativistic
behavior. Similarly, for $c_{\perp}\rightarrow0$, the emergent Lorentz
symmetry break down. This means that $\Lambda$ or $\beta^{-1/2}$
for the last two terms in Eq. (\ref{eq:momentum_anomaly_susy}) is
$mc_{\perp}^{2}$. In $^{3}\text{He-A}$, it is known that $c_{\perp}^{2}m^{2}/p_{F}^{2}\simeq10^{-5}$,
so the last two terms in Eq. (\ref{eq:momentum_anomaly_susy}) is
negligible compared to the first term in Eq. (\ref{eq:momentum_anomaly_susy}).
Especially, the term in the third line in Eq. (\ref{eq:momentum_anomaly_susy})
is the Nieh-Yan anomaly. Hence, we have shown that the Nieh-Yan term
is negligible in the $^{3}\text{He-A}$ phase. 

\section{Conclusion \label{sec:conclusion}}

We have employed the supersymmetric quantum mechanics to derive both
the (chiral) $U\left(1\right)$-gauge anomaly and the (chiral) diffeomorphism
anomaly in the $^{3}\text{He-A}$ phase, the chiral superconductors
and the Weyl semimetals. In contrast to most known anomalies, the
scale from the position of Weyl nodes in the energy-momentum space
can enter the anomaly equations, whose physical origin is clarified
from the view of spectral flow. The bulk-dislocation correspondence
is then constructed based on these anomaly equations. For systems
with particel-hole symmetry, these trapped zero modes should be Majorana
Fermions. In addition, the momentum anomaly in the $^{3}\text{He-A}$
phase naturally arises from our diffeomorphism anomaly equation. The
well-established results in the $^{3}\text{He-A}$ phase enable us
to fix the ambiguities in the torsional anomalies from cut-off dependence.
Because the Lorentz symmetry breaking scale is relatively small compared
to the chemical potential, the Nieh-Yan term as well as other cut-off
dependence terms are shown to be negligible. 

In addition to dislocations, strain can cause lattice deformations
as well. That is, the effective model for the strained Weyl system
is given in Eq. (\ref{eq:action}). If the separation between Weyl
nodes equal to the reciprocal vector, then we have obtained the weak
topological insulators with the stacking direction along $\lambda_{a}$.
Hence, our results should be applicable to the strained Weyl systems
as well as the weak topological insulators. This suggests that the
chiral zero modes can be manipulated by strains.

\section*{Acknowledgement}

The authors thank Aris Alexandradninata, Jaakko Nissinen and Mike
Stone for insightful discussions, especially Mike for invaluable feedback.
The authors greatly appreciate Onkar Parrikar for explaining the technical
details in Ref. \citep{parrikar2014prd}. Z.-M. H were not directly
supported by any funding agency, but this work would not be possible
without resources provided by the Department of Physics at the University
of Illinois at Urbana-Champaign. B. H. was supported by ERC Starting
Grant No. 678795 TopInSy.

\appendix

\section{Torsional Schrodinger-Lichnerowicz identity \label{sec:torsional_Lichnerowi}}

In this section, we shall derive the torsional version of the Schrodinger-Lichnerowicz
identity. For simplicity, we shall focus on the $\left(2n\right)$-dim
Euclidean spacetime in this section and Append. \ref{sec:susy_PI}.
The Dirac operator is $\slashed{D}=e_{a}^{\mu}\gamma^{a}\left(\partial_{\mu}+\frac{1}{2}\omega_{ab\mu}\sigma^{ab}\right)$,
where $a=1,\ 2,\ \dots,\ \left(2n\right)$ and $\gamma^{a}$ in the
Euclidean spacetime satisfies $\left\{ \gamma^{a},\ \gamma^{b}\right\} =2\delta^{ab}$.
The $U\left(1\right)$-gauge field is temporarily tuned to zero. However,
in the presence of torsion, $\slashed{D}$ defined here is not skew-Hermitian,
i.e., 
\[
\langle\bar{\psi}|\slashed{D}\varphi\rangle\neq-\langle\left(\slashed{D}\bar{\psi}\right)|\varphi\rangle,
\]
where $\langle\bar{\psi}|\varphi\rangle\equiv\int d^{d}x\sqrt{\left|g\right|}\bar{\psi}\left(x\right)\varphi\left(x\right)$.
Notice that $\slashed{D}$ can be rendered to be skew Hermitian by
adding an extra piece, i.e., $\frac{1}{2}{T^{\rho}}_{\mu\rho}$, where
${T^{\rho}}_{\mu\nu}$ is the torsion tensor. Hence, we can define
\begin{eqnarray*}
\mathcal{D}_{\mu} & = & \partial_{\mu}+\frac{1}{2}\omega_{ab\mu}\sigma^{ab}+\frac{1}{2}{T^{\rho}}_{\mu\rho},
\end{eqnarray*}
and 
\begin{eqnarray*}
\overleftarrow{\slashed{\mathcal{D}}} & = & \left(\overleftarrow{\partial}_{\mu}-\frac{1}{2}\omega_{ab\mu}\sigma^{ab}+\frac{1}{2}{T^{\rho}}_{\mu\rho}\right)e_{a}^{\mu},
\end{eqnarray*}
which are now skew Hermitian, i.e., 
\[
\langle\bar{\psi}|\slashed{\mathcal{D}}\varphi\rangle=-\langle\slashed{\mathcal{D}}\bar{\psi}|\varphi\rangle.
\]
 In terms of $\slashed{\mathcal{D}}$ and $\overleftarrow{\slashed{\mathcal{D}}}$,
the action is intact, i.e., 
\begin{eqnarray*}
 &  & \frac{1}{2}\int d^{d}x\sqrt{\left|g\right|}\left(\bar{\psi}\mathcal{\slashed{\mathcal{D}}\psi}-\bar{\psi}\overleftarrow{\slashed{\mathcal{D}}}\psi\right)\\
 & = & \frac{1}{2}\int d^{d}x\sqrt{\left|g\right|}\left(\bar{\psi}\slashed{D}\psi-\bar{\psi}\overleftarrow{\slashed{D}}\psi\right).
\end{eqnarray*}
For chiral transformation $\psi\rightarrow e^{i\theta\gamma^{2n+1}}\psi$,
the corresponding Jacobian is known to be \citep{fujikawa2004oxford,bertlmann2000oxford}
\[
\ln J\left(\theta\right)=-2i\theta\text{Tr}\gamma^{2n+1},
\]
which is divergent. The regularized Jacobian is 
\[
\ln J\left(\theta\right)=-2i\theta\lim_{\beta\rightarrow0}\text{Tr}\gamma^{2n+1}e^{\beta\slashed{\mathcal{D}}\slashed{\mathcal{D}}}.
\]
The rest part of this section is devoted to derive the Schrodinger-Lichnerowicz
identity for $\slashed{\mathcal{D}}\slashed{\mathcal{D}}$. Before
diving into lengthy derivations, we shall list our conventions here:
\begin{enumerate}
\item $\mathring{}$ is used for the torsion-free quantities, for example,
$\mathring{D}_{\mu}\equiv\partial_{\mu}+\frac{1}{2}\mathring{\omega}_{ab\mu}\sigma^{ab}$.
$\mathring{\omega}_{ab\mu}$ is the spin connection satisfies the
torsion-free condition and ${\mathring{\Gamma}_{\ }^{\lambda}}{}_{\mu\nu}$
is the Christoffel connection, i.e., ${\mathring{\Gamma}_{\ }^{\lambda}}{}_{\mu\nu}=\frac{1}{2}g^{\lambda\rho}\left(-\partial_{\rho}g_{\mu\nu}+\partial_{\mu}g_{\rho\nu}+\partial_{\nu}g_{\rho\mu}\right)$;
\item $\mathcal{D}_{\mu}\equiv\partial_{\mu}+\frac{1}{2}\omega_{ab\mu}\sigma^{ab}+\frac{1}{2}{T^{\rho}}_{\mu\rho}=\partial_{\mu}+\frac{1}{2}\left(\mathring{\omega}_{\mu ab}-\frac{1}{6}H_{\mu ab}\right)\sigma^{ab}$,
where $H_{abc}$ is a totally anti-symmetric tensor, i.e., $H_{abc}=3T_{\left[abc\right]}$
and $T^{a}=de^{*a}+{\omega^{a}}_{b}\wedge e^{*b}$;
\item $\widehat{\mathcal{D}}_{\mu}\equiv\partial_{\mu}+\frac{1}{2}\left(\mathring{\omega}_{\mu ab}-\frac{1}{2}H_{\mu ab}\right)\sigma^{ab}$
and the corresponding curvature is written as $\widehat{\Omega}_{ab}=d\widehat{\omega}_{ab}+\left(\widehat{\omega}\wedge\widehat{\omega}\right)_{ab}$;
\item We used $D_{\mu}$ for the (Lorentz) covariant derivative, whose connection
acts on the Lorentz indices. $\nabla_{\mu}$ for the totally covariant
derivative, whose connection acts on both the Lorentz and the Einstein
indices. The same definition holds for $\widehat{\mathcal{D}}_{\mu}$,
$\mathring{D}_{\mu}$ and $\widehat{\nabla}_{\mu}$, $\mathring{\nabla}_{\mu}$;
\item The contorsion $C_{ab\mu}$ is defined as $\omega_{ab\mu}=\mathring{\omega}_{ab\mu}+C_{ab\mu}$,
or $C_{abc}=-\frac{1}{2}\left(T_{abc}-T_{cab}+T_{bca}\right)$. 
\end{enumerate}
The torsional Schrodinger-Lichnerowicz identity for $\slashed{\mathcal{D}}\slashed{\mathcal{D}}$
is given as \citep{peeters1999jhep}

\begin{eqnarray}
 &  & \gamma^{\mu}\mathcal{D}_{\mu}\gamma^{\nu}\mathcal{D}_{\nu}\nonumber \\
 & = & g^{\mu\nu}\widehat{\mathcal{D}}_{\mu}\widehat{\mathcal{D}}_{\nu}-g^{\mu\lambda}{\Gamma^{\nu}}_{\mu\lambda}\widehat{\mathcal{D}}_{\nu}+\frac{1}{6}g^{\mu\nu}\left(\mathring{\nabla}_{\mu}H_{\nu cd}\right)\sigma^{cd}\nonumber \\
 &  & +\frac{1}{2}\sigma^{ab}\sigma^{cd}\left[\Omega_{cd,\ ab}-2\left(\frac{1}{6}\right)^{2}H_{\mu ab}H_{\nu cd}g^{\mu\nu}\right],\label{eq:Lichnerowicz}
\end{eqnarray}
and the torsional Schrodinger-Lichnerowicz identity for the operator
with both $U\left(1\right)$-gauge fields and $\lambda_{a}$ is presented
in Eq. (\ref{eq:Lichnerowicz-1}). 

We are now ready to present the derivation of Eq. (\ref{eq:Lichnerowicz}).

Because of the vielbein postulate, there is
\[
\left(\partial_{\mu}+\frac{1}{2}\omega_{ab\mu}\sigma^{ab}\right)\gamma^{\nu}=\left[\gamma^{\nu}\left(\partial_{\mu}+\frac{1}{2}\omega_{ab\mu}\sigma^{ab}\right)-\gamma^{\lambda}{\Gamma^{\nu}}_{\lambda\mu}\right].
\]
This implies that 

\begin{eqnarray}
 &  & \gamma^{\mu}\mathcal{D}_{\mu}\gamma^{\nu}\mathcal{D}_{\nu}\nonumber \\
 & = & \left[\gamma^{\mu}\gamma^{\nu}\mathcal{D}_{\mu}\left(\omega\right)-\gamma^{\mu}\gamma^{\lambda}{\Gamma^{\nu}}_{\lambda\mu}\right]\mathcal{D}_{\nu}\nonumber \\
 & = & \left(g^{\mu\nu}\mathcal{D}_{\mu}\mathcal{D}_{\nu}-g^{\mu\lambda}{\Gamma^{\nu}}_{\mu\lambda}\mathcal{D}_{\nu}\right)+\nonumber \\
 &  & +\sigma^{\mu\nu}\left[\mathcal{D}_{\mu},\ \mathcal{D}_{\nu}\right]+2g^{\mu\nu}\sigma^{\beta\sigma}C_{\mu\beta\sigma}\mathcal{D}_{\nu},\label{eq:D^2}
\end{eqnarray}
where in the last line, we have used 
\[
C_{\alpha\beta\sigma}\sigma^{\beta\sigma}=-\frac{1}{2}\left(T_{\alpha\beta\sigma}-T_{\sigma\alpha\beta}+T_{\beta\sigma\alpha}\right)\sigma^{\beta\sigma}=-\frac{1}{2}\sigma^{\beta\sigma}T_{\alpha\beta\sigma},
\]
and the last equality is due to the antisymmetric properties of $\sigma^{\beta\sigma}$,
while $-T_{\sigma\alpha\beta}+T_{\beta\sigma\alpha}$ is symmetric
upon the indices $\beta,\ \sigma$. By definition, there is $C_{\alpha\beta\sigma}=-C_{\beta\alpha\sigma}$,
so $C_{a\beta\sigma}\sigma^{\beta\sigma}=C_{\left[\alpha\beta\sigma\right]}\sigma^{\beta\sigma}$.
That is, only the totally antisymmetric components are non-zero in
the term $C_{\alpha\beta\sigma}\sigma^{\beta\sigma}$, so there is
\[
C_{\alpha\beta\sigma}\sigma^{\beta\sigma}=-\frac{1}{6}H_{\alpha bc}\sigma^{bc},
\]
where the identity $T_{\left[abc\right]}=\frac{1}{3}H_{abc}$ is used.
Consequently, Eq. (\ref{eq:D^2}) can be further recast as 

\begin{eqnarray}
 &  & \slashed{\mathcal{D}}^{2}=\left(g^{\mu\nu}\mathcal{D}_{\mu}\mathcal{D}_{\nu}-g^{\mu\lambda}{\Gamma^{\nu}}_{\mu\lambda}\mathcal{D}_{\nu}\right)\nonumber \\
 &  & +\sigma^{\mu\nu}\left[\mathcal{D}_{\mu},\ \mathcal{D}_{\nu}\right]-4\frac{1}{12}g^{\mu\nu}\sigma^{\beta\sigma}H_{\mu\beta\sigma}\mathcal{D}_{\nu}.\label{eq:D^2_H}
\end{eqnarray}
Now we want to rewrite the equation above in terms of $\widehat{\mathcal{D}}_{\mu}=\mathcal{D}_{\mu}+\left(-\frac{1}{6}H_{\mu ab}\sigma^{ab}\right).$
This can be done by recast $\mathcal{D}_{\mu}$ in terms of $\widehat{\mathcal{D}}_{\mu}$
and $-\frac{1}{6}H_{\mu ab}\sigma^{ab}$. Then, after some lengthy
algebraic manipulations, one can obtain Eq. (\ref{eq:Lichnerowicz}).
Especially, for the following operator 
\[
\gamma^{a}e_{a}^{\mu}\left(\partial_{\mu}+\frac{1}{2}\omega_{ab\mu}\sigma^{ab}+iA_{\mu}\right)+i\lambda_{a}\gamma^{a},
\]
the corresponding Shrodinger-Lichnerowicz can be derived in a parallel
manner, i.e., 

\begin{eqnarray}
 &  & \left(\gamma^{\mu}\mathcal{D}_{\mu}+i\lambda_{a}\gamma^{a}\right)\left(\gamma^{\nu}\mathcal{D}_{\nu}+i\lambda_{a}\gamma^{a}\right)\nonumber \\
 & = & g^{\mu\nu}\widehat{\mathcal{D}}_{\mu}\widehat{\mathcal{D}}_{\nu}-g^{\mu\lambda}{\Gamma^{\nu}}_{\mu\lambda}\widehat{\mathcal{D}}_{\nu}+\frac{1}{6}g^{\mu\nu}\left(\mathring{\nabla}_{\mu}H_{\nu cd}\right)\sigma^{cd}\nonumber \\
 &  & +i\left(F_{ab}+\lambda_{c}{\tilde{T}_{\ }^{c}}{}_{ab}\right)\sigma^{ab}\nonumber \\
 &  & +\frac{1}{2}\sigma^{ab}\sigma^{cd}\left[\Omega_{cdab}-2\left(\frac{1}{6}\right)^{2}H_{\mu ab}H_{\nu cd}g^{\mu\nu}\right],\label{eq:Lichnerowicz-1}
\end{eqnarray}
where ${\tilde{T}_{\ }^{a}}{}_{\mu\nu}$ is defined as ${\tilde{T}_{\ }^{a}}{}_{\mu\nu}=\partial_{\mu}e_{\nu}^{*a}-\partial_{\nu}e_{\mu}^{*a}$
and it is different from the torsion by the spin connection term. 

\section{Supersymmetric quantum mechanics and Chiral anomaly \label{sec:susy_PI}}
\begin{widetext}
In this section, we shall first represent the Jacobian of chiral anomaly
in terms of the supersymmetric quantum mechanics. The receipt is presented
in details in Ref. \citep{boer1996npb,bastianelli2006cambridge},
so we shall only outline the main steps here. After that, we shall
calculate this Jacobian by perturbative calculations. 

\begin{figure}
\begin{widetext}
\includegraphics[scale=0.5]{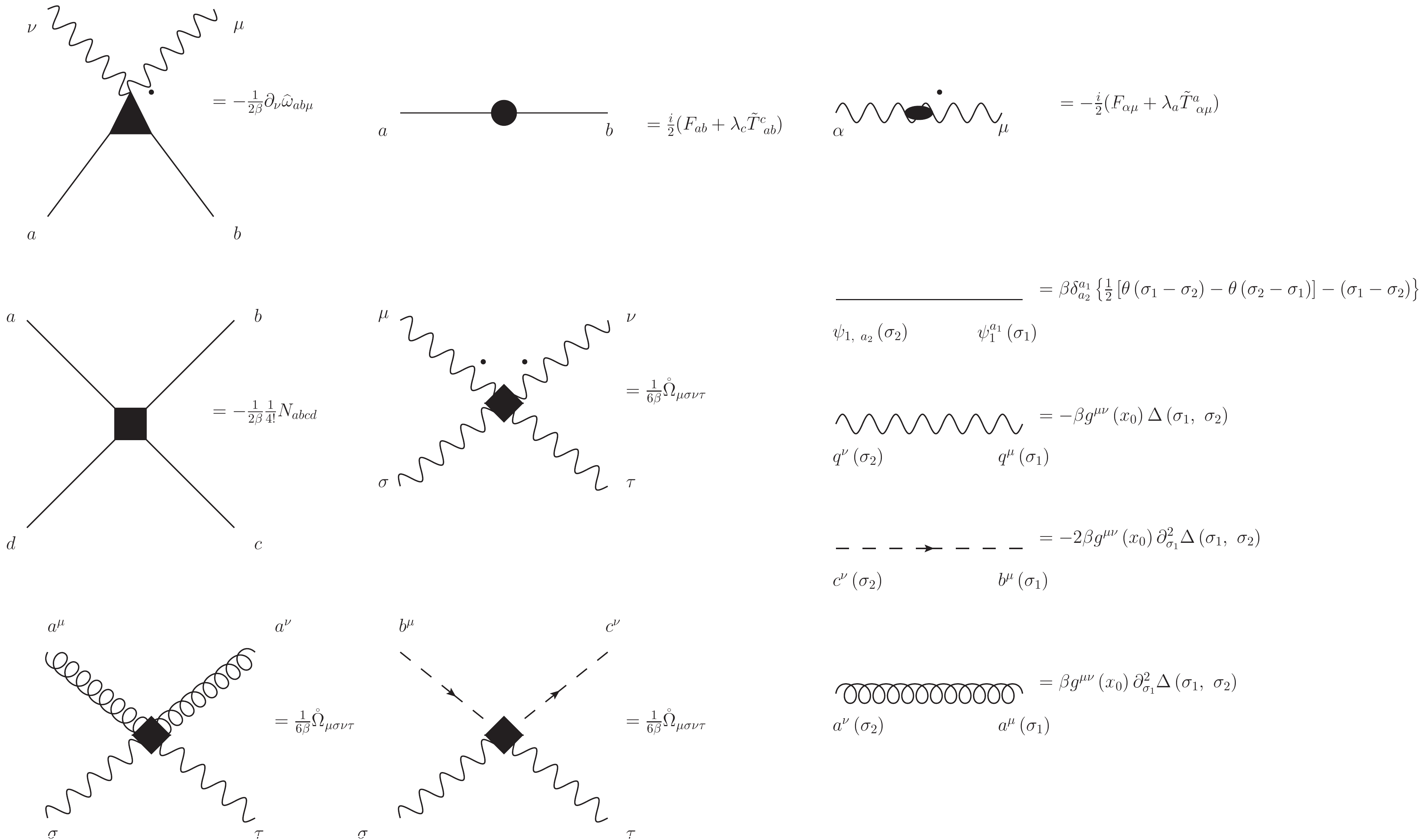}

\caption{Feynman rules for the Jacobian represented by the supersymmetric quantum
mechanics \label{fig:Feynman_rules}. $\Delta\left(\sigma_{1},\ \sigma_{2}\right)$
is defined as $\Delta\left(\sigma_{1},\ \sigma_{2}\right)=\sigma_{1}\left(\sigma_{2}+1\right)\theta\left(\sigma_{1}-\sigma_{2}\right)+\sigma_{2}\left(\sigma_{1}+1\right)\theta\left(\sigma_{2}-\sigma_{1}\right)$.
For the Fermionic fields, they can be either the background fields
$\psi_{1,\ \text{bg}}^{a}$ or the fluctuating fields, $\psi_{1,\ \text{qu}}^{a}$.
We have not distinguished them in the Feynman rules here. In the Feynman
diagrams shown later, the background Fermionic fields shall be labeled
by connecting to a cross symbols. }
\end{widetext}

\end{figure}
\end{widetext}

\subsection{Supersymmetric quantum mechanics and the Jacobian for chiral anomaly}
\begin{widetext}
The supersymmetric quantum mechanics representation of the chiral
anomaly Jacobian is
\[
J\left(\theta\right)=-2i\theta\lim_{\beta\rightarrow0}\text{Tr}\gamma^{2n+1}e^{\beta\slashed{\mathcal{D}}\slashed{\mathcal{D}}},
\]
which reminds us of the transition amplitude in the Euclidean spacetime,
i.e., $\langle\bar{\psi}|e^{-TH}|\varphi\rangle$ with $H=-\slashed{\mathcal{D}}\slashed{\mathcal{D}}$.
Notice that the gamma matrices satisfy $\left\{ \gamma^{a},\ \gamma^{b}\right\} =2\delta^{ab}$
and the Majorana Fermion $\hat{\psi}_{1}^{a}$ satisfies $\left\{ \hat{\psi}_{1}^{a},\ \hat{\psi}_{1}^{b}\right\} =\delta^{ab}$,
so we can represent the gamma matrix $\gamma^{a}$ in terms of the
Majorana Fermion, i.e., $\gamma^{a}=\sqrt{2}\hat{\psi}^{a}$. In
addition, for later convenience, we shall adopt the Riemann normal
coordinate, i.e., 
\begin{equation}
g_{\mu\nu}=\delta_{\mu\nu}-\frac{1}{3}\mathring{\Omega}_{\mu\sigma,\ \nu\tau}q^{\sigma}q^{\tau}+\mathcal{O}\left(\left|q\right|^{3}\right),\label{eq:metric_RiemannNormal}
\end{equation}
or 
\[
\partial_{\alpha}\partial_{\beta}g_{\mu\nu}=-\frac{2}{3}\mathring{\Omega}_{\mu(\alpha,\ |\nu|\beta)}+\mathcal{O}\left(\left|q\right|\right),
\]
and 
\begin{equation}
{\mathring{\Gamma}_{\ }^{\lambda}}{}_{\mu\nu}=-\frac{1}{3}\left({\mathring{\Omega}_{\ }^{\lambda}}{}_{\mu\nu\tau}+{\mathring{\Omega}_{\ }^{\lambda}}{}_{\nu\mu\tau}\right)q^{\tau}+\mathcal{O}\left(\left|q\right|^{2}\right).\label{eq:Christoff_RiemannNormal}
\end{equation}
If we choose such a frame field that $e_{\mu}^{*a}=\delta_{\mu}^{a}-\frac{1}{6}{\mathring{\Omega}_{\ }^{a}}{}_{\sigma\mu\tau}q^{\sigma}q^{\tau},$
then one can recover the metric in Eq. (\ref{eq:metric_RiemannNormal})
and the spin connection is given as ${\mathring{\omega}_{\ }^{a}}{}_{b\mu}=-e_{b}^{\nu}\left(\partial_{\mu}e_{\nu}^{a}-{\mathring{\Gamma}^{a}}{}_{\nu\mu}\right).$
Due to the local Lorentz symmetry, we can perform such a transformation
of $e_{\mu}^{*a}$ that $\widehat{\omega}_{ab\mu}=0$, i.e., $e_{\mu}^{*a}\rightarrow{R^{a}}_{b}e_{\mu}^{*b}$
and $\widehat{\omega}_{ab\mu}\rightarrow R^{-1}\widehat{\omega}_{ab\mu}R-\left(R^{-1}DR\right)_{ab}=0$.
That is, we can choose the Riemann normal coordinate as well as such
a frame that $\widehat{\omega}_{\mu}=0$. 

By using the method of background fields, the Jacobian can be recast
as

\begin{equation}
J/\left(-2i\theta\right)=\left(\frac{-i}{2\pi}\right)^{d/2}\int d^{d}x\sqrt{\left|g\right|}\int\left(\prod_{a=1}^{d}d\psi_{\text{bg}}^{a}\right)\langle e^{-S_{\text{int}}}\rangle,
\end{equation}
where $\psi_{\text{bg}}^{a}$ is the background Fermionic field and
$S_{\text{int}}$ is given as 
\begin{eqnarray}
S_{\text{int}} & = & \frac{1}{2\beta}\int_{-1}^{0}d\sigma\left(\frac{d}{d\sigma}q^{\mu}\frac{d}{d\sigma}q^{\nu}+b^{\mu}c^{\nu}+a^{\mu}a^{\nu}\right)\left(-\frac{1}{3}\mathring{\Omega}_{\mu\sigma,\ \nu\tau}q^{\sigma}q^{\tau}\right)\nonumber \\
 &  & +\frac{i}{2}\int_{-1}^{0}d\sigma\left(F_{\alpha\mu}+\lambda_{c}{\tilde{T}_{\ }^{c}}{}_{\alpha\mu}\right)\frac{d}{d\sigma}q^{\mu}q^{\alpha}\nonumber \\
 &  & +\frac{1}{2\beta}\int_{-1}^{0}d\sigma\partial_{\nu}\widehat{\omega}_{ab\mu}\psi_{1}^{a}\psi_{1}^{b}q^{\nu}\frac{d}{d\sigma}q^{\mu}\nonumber \\
 &  & +\frac{1}{2\beta}\int_{-1}^{0}d\sigma\frac{1}{4!}N_{abcd}\psi_{1}^{a}\psi_{1}^{b}\psi_{1}^{c}\psi_{1}^{d}\nonumber \\
 &  & -\frac{i}{2}\int_{-1}^{0}d\sigma\left(F_{ab}+\lambda_{c}{\tilde{T}_{\ }^{c}}{}_{ab}\right)\psi_{1}^{a}\psi_{1}^{b}\nonumber \\
 &  & +\frac{\beta}{8}\int_{-1}^{0}d\sigma\left[3\left(\frac{1}{6}\right)^{2}\text{tr}H_{\mu}H_{\nu}\right]g^{\mu\nu}\left(x_{0}\right).\label{eq:action_RC_Frame}
\end{eqnarray}
The fields $b^{\mu}$ and $c^{\nu}$ are anti-commutating, while $a^{\mu}$
is commutating. They are from the $\det g_{\mu\nu}$ in the path-integral
quantization. $N_{abcd}$ is the Nieh-Yan term, i.e., $N=dH$. The
last term in $S_{\text{int}}$, $\frac{\beta}{8}\int_{-1}^{0}d\sigma\left[3\left(\frac{1}{6}\right)^{2}\text{tr}H_{\mu}H_{\nu}\right]g^{\mu\nu}\left(x_{0}\right)$,
is the so-called counter terms and they are from the Weyl ordering.
The corresponding Feynman rules are shown in Fig. \ref{fig:Feynman_rules}.
Note that the Fermionic field $\psi^{a}_{1}$ satisfies periodic boundary
condition and this is the origin of the extra piece $-\left(\sigma_{1}-\sigma_{2}\right)$
in the propagator \footnote{We thank Onkar Parrikar for pointing out this.}. 

\begin{figure}
\includegraphics[scale=0.5]{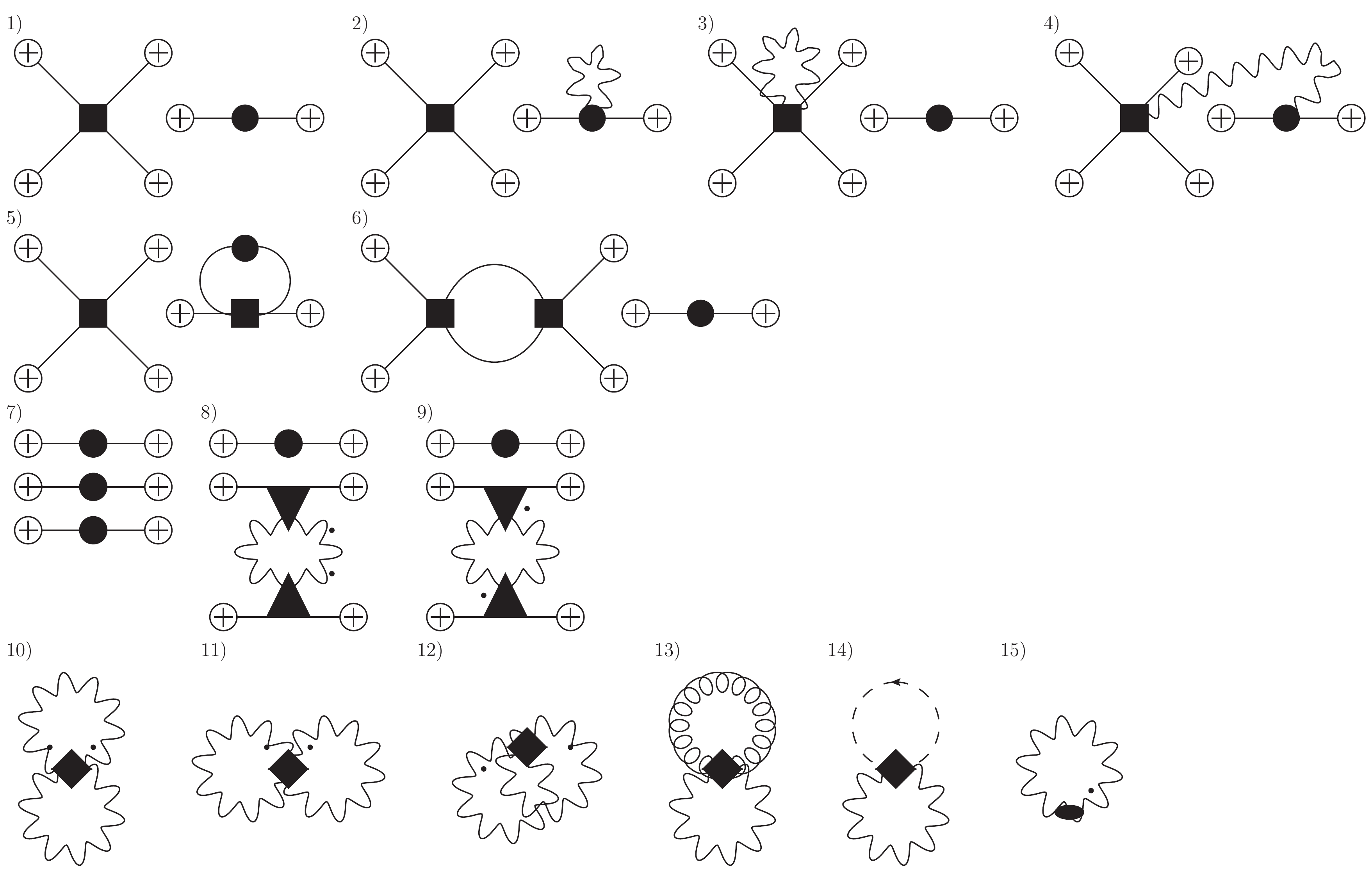}\caption{Non-zero Feynman diagrams \label{fig:Nonzero_FeynmanDiagrams}}

\end{figure}
\end{widetext}

\subsection{Perturbative calculations of the Jacobian }

Because $\beta$ is sent to infinity, the Jacobian can be calculated
by perturbative calculations. All the Feynman diagrams can be organized
by counting the power of $\beta$. For clarity, let us first consider
a Feynman diagram consisting only of vertices with Fermionic fields,
for example, a Feynman diagrams with $y$ Nieh-Yan vertices, $n$
$\left(F_{ab}+\lambda_{c}{\tilde{T}_{\ }^{c}}{}_{ab}\right)\psi_{1}^{a}\psi_{1}^{b}$
vertices and $m$ $\partial_{\nu}\widehat{\omega}_{ab\mu}\psi^{a}\psi^{b}q^{\nu}\frac{d}{d\sigma}q^{\mu}$
vertices. Because in $6$-dim spacetime, there must exist $6$ external
background Fermionic fields so as to ensure the integral $\int\left(\prod_{a=1}^{d}d\psi_{\text{bg}}^{a}\right)$
non-zero, the order $\beta$ of this Feynman diagram is 
\[
\mathbf{\beta}^{y+n+m-3}.
\]
Since the order of $\beta$ for a given Feynman diagrams is required
to be non-positive in the limit $\beta\rightarrow0$, we can determine
the value of $y,\ n$ and $m$ by requiring 
\begin{equation}
y+n+m-3\leq0.
\end{equation}
For $y=0$, there are
\begin{enumerate}
\item $n=3,\ m=0,$ $\beta^{0}$;
\item $n=2,\ m=1$, $\beta^{0}$;
\item $n=1,\ m=2$, $\beta^{0}$;
\item $n=0,\ m=3$, $\beta^{0}$, 
\end{enumerate}
For $y=1$, there are
\begin{enumerate}
\item $n=2,\ m=0$, $\beta^{0}$;
\item $n=1,\ m=0$, $\beta^{-1}$;
\item $n=1,\ m=1$, $\beta^{0}$
\item $n=0,\ m=1$, $\beta^{-1}$;
\item $n=0,\ m=2$, $\beta^{0}$;
\end{enumerate}
For $y=2$, there are
\begin{enumerate}
\item $n=1,\ m=0$, $\beta^{0}$; 
\item $n=0,\ m=0,\ \beta^{-1}$; 
\item $n=0,\ m=1$, $\beta^{0}$;
\end{enumerate}
For $y=3$, there is $n=m=0$. 

These diagrams are dubbed as the skeleton diagrams. As for the vertices
in Fig. \ref{fig:Feynman_rules} without Fermionic fields, they are
of order $\mathcal{O}\left(\beta\right)$. For example, the order
of $-\frac{1}{2}F_{\alpha\mu}q^{\alpha}\frac{d}{d\sigma}q^{\mu}$
is $\mathcal{O}\left(\beta\right)$, where $q^{\alpha}$ and $\frac{d}{d\sigma}q^{\mu}$
are both of order $\mathcal{O}\left(\beta^{1/2}\right)$. Hence, these
vertices can only appear with the skeleton diagrams with order $\mathcal{O}\left(\beta^{-1}\right)$.
All the non-zero Feynman diagrams are shown in Fig. \ref{fig:Nonzero_FeynmanDiagrams}. 

\subsection{Six dimensional chiral anomaly}
\begin{widetext}
By calculating the diagrams listed in Fig. \ref{fig:Nonzero_FeynmanDiagrams},
one can obtain 

\begin{eqnarray}
 &  & J/\left(-2i\theta\right)\nonumber \\
 & = & \frac{1}{8\pi^{3}}\int\{-\frac{1}{2\beta}N\wedge\left(F+\lambda_{a}\tilde{T}^{a}\right)-\frac{1}{6}\left(F+\lambda_{a}\tilde{T}^{a}\right)\wedge\left(F+\lambda_{a}\tilde{T}^{a}\right)\wedge\left(F+\lambda_{a}\tilde{T}^{a}\right)+\frac{1}{48}\left(F+\lambda_{a}\tilde{T}^{a}\right)\wedge\widehat{\Omega}_{\mu\nu}^{-}\wedge\widehat{\Omega}_{\mu\nu}^{-}\nonumber \\
 &  & -\frac{1}{24}\partial^{2}\left[N\wedge\left(F+\lambda_{a}\tilde{T}^{a}\right)\right]+\frac{1}{24}\partial N\wedge\partial\left(F+\lambda_{a}\tilde{T}^{a}\right)\nonumber \\
 &  & -\frac{1}{48}N\wedge\left[\left(F_{\mu\nu}+\lambda_{a}{\tilde{T}_{\ }^{a}}{}_{\mu\nu}\right)\widehat{\Omega}_{\mu\nu}^{-}\right]+\frac{1}{48}\widehat{\Omega}^{-}N\wedge\left(F+\lambda_{a}\tilde{T}^{a}\right)-\frac{1}{48}\left(F+\lambda_{a}\tilde{T}^{a}\right)\wedge N_{\mu\nu}\wedge\widehat{\Omega}_{\mu\nu}^{-}\},\label{eq:Jacobian_1}
\end{eqnarray}
where $H_{ab}\equiv i_{e_{b}}i_{e_{a}}H$, $\widehat{\omega}_{ab}^{-}\equiv\mathring{\omega}_{ab}+\frac{1}{2}H_{ab}$,
$\widehat{\Omega}_{ab,\ \mu\nu}^{-}\equiv i_{\partial_{\nu}}i_{\partial_{\mu}}\left[d\widehat{\omega}_{ab}^{-}+\left(\widehat{\omega}^{-}\wedge\widehat{\omega}^{-}\right)_{ab}\right]$,
$\widehat{\Omega}_{\mu\nu}^{-}\equiv\frac{1}{2}\widehat{\Omega}_{ab,\ \mu\nu}^{-}e^{*a}\wedge e^{*b}$
and $\widehat{\Omega}^{-}$ is the Ricci scalar associated with $\widehat{\Omega}_{\rho\sigma,\ \mu\nu}^{-}$.
In the Riemann normal coordinate, there are $\widehat{\Omega}_{\mu\nu,\ \rho\sigma}^{-}=\widehat{\Omega}_{\rho\sigma,\ \mu\nu}^{-}$
and thus $\widehat{\Omega}_{\mu\nu}^{-}=\frac{1}{2}\widehat{\Omega}_{\mu\nu,\ \rho\sigma}^{-}dx^{\rho}\wedge dx^{\sigma}$. 

Because $dF=d\tilde{T}^{a}=dN=0$, the last term in the second line
of Eq. (\ref{eq:Jacobian_1}) can be written as 

\[
\partial N\wedge\partial\left(F+\lambda_{a}\tilde{T}^{a}\right)=dN_{m}\wedge d\left(F_{n}+\lambda_{a}{\tilde{T}_{\ }^{a}}{}_{n}\right)\eta^{mn},
\]
where $N_{m}\equiv i_{e_{m}}N$, $F_{m}\equiv i_{e_{m}}F$ and ${\tilde{T}_{\ }^{a}}{}_{n}\equiv i_{e_{n}}\tilde{T}^{a}$.
As for terms in the last line in Eq. (\ref{eq:Jacobian_1}), they
can be recast as 

\begin{eqnarray}
 &  & -\frac{1}{48}N\wedge\left[\left(F_{\mu\nu}+\lambda_{a}{\tilde{T}_{\ }^{a}}{}_{\mu\nu}\right)\widehat{\Omega}_{\mu\nu}^{-}\right]+\frac{1}{48}\widehat{\Omega}^{-}N\wedge\left(F+\lambda_{a}\tilde{T}^{a}\right)-\frac{1}{48}\left(F+\lambda_{a}\tilde{T}^{a}\right)\wedge N_{\mu\nu}\wedge\widehat{\Omega}_{\mu\nu}^{-}\nonumber \\
 & = & -\frac{i}{24}\widehat{\Omega}_{ab}^{-}\wedge N_{a}\wedge\left(F_{b}+\lambda_{c}{\tilde{T}_{\ }^{c}}{}_{b}\right).\label{eq:hodge_dual}
\end{eqnarray}
This can be derived by rewriting the Nieh-Yan term $N$ in terms of
its Hodge dual $\tilde{N}=\star N$, i.e., $N=\star\tilde{N}$. Then,
by calculating the contraction between the two Levi-Civita antisymmetric
tensor (one from the volume form and one from the Hodge dual), one
can obtain Eq. (\ref{eq:hodge_dual}).

Finally, after these algebraic manipulations, the Jacobian is given
as 

\begin{eqnarray}
 &  & J/\left(-2i\theta\right)\nonumber \\
 & = & \frac{1}{8\pi^{3}}\int\{-\frac{1}{2\beta}N\wedge\left(F+\lambda_{a}\tilde{T}^{a}\right)\nonumber \\
 &  & -\frac{1}{6}F\wedge F\wedge F-\frac{1}{48}\left(F+\lambda_{a}\tilde{T}^{a}\right)\wedge\text{tr}\left(\widehat{\Omega}^{-}\wedge\widehat{\Omega}^{-}\right)\nonumber \\
 &  & -\frac{1}{24}d\star d\star\left[N\wedge\left(F+\lambda_{a}\tilde{T}^{a}\right)\right]+\frac{1}{24}dN\wedge d\left(F+\lambda_{a}\tilde{T}^{a}\right)\nonumber \\
 &  & -\frac{1}{24}\widehat{\Omega}_{ab}^{-}\wedge N_{a}\wedge\left(F_{b}+\lambda_{a}{\tilde{T}_{\ }^{a}}{}_{b}\right)\},\label{eq:jacobian_chiralanomaly}
\end{eqnarray}
which matches with the previous results if we set $\lambda_{a}=0$
\footnote{O. Parrikar, private communication}. Since we are most
interested in the adiabatic limit, terms in the $4$-th line shall
be neglected, because they have higher-order derivative. The term
in the last line shall be neglected in the $\left(4+1\right)$-dim
parity-odd effective action as well, because that it can not be recast
into a surface term. 
\end{widetext}

\section{Covariant Lie derivative, Killing vector and conservation law \label{sec:EM_conservation}}
\begin{widetext}
In this section, we shall derive the energy-momentum conservation
law associated with the covariant Lie derivative. Then, by defining
the Killing vectors properly, the conservation law can be recast to
a rather illustrating form. 

The covariant Lie derivative is defined as 
\[
\delta_{\xi}^{C}e_{\nu}^{*a}=\xi^{\mu}{T^{a}}_{\mu\nu}+\nabla_{\nu}\xi^{a},
\]
\[
\delta_{\xi}^{C}\omega_{ab\nu}=\Omega_{ab,\ \mu\nu}\xi^{\mu},
\]
and 
\begin{eqnarray*}
\delta_{\xi}^{C}A_{\nu} & = & F_{\mu\nu}\xi^{\mu}.
\end{eqnarray*}
This transformation is obtained by performing a Lie derivative associated
with the vector $\xi$, then a rotation $i_{\xi}\omega$ and finally
a $U\left(1\right)$ transformation $\exp\left[i\left(\xi^{\mu}A_{\mu}\right)\right]$. 

The variation of the action $S\left(e_{\mu}^{*a},\ \omega,\ A\right)$
is 

\begin{eqnarray*}
\delta S & = & \int d^{d}x\sqrt{\left|g\right|}\left(\frac{1}{\sqrt{\left|g\right|}}\frac{\delta S}{\delta e_{\nu}^{*a}}\delta e_{\nu}^{*a}+\frac{1}{\sqrt{\left|g\right|}}\frac{\delta S}{\delta\omega_{ab\nu}}\delta\omega_{ab\nu}+\frac{1}{\sqrt{\left|g\right|}}\frac{\delta S}{\delta A_{\nu}}\delta A_{\nu}\right)\\
 & = & \int d^{d}x\sqrt{\left|g\right|}\xi^{a}\left[\left(\nabla_{\nu}+{T^{\rho}}_{\nu\rho}\right)\tau_{a}^{\nu}-e_{a}^{\mu}\left(\tau_{a}^{\nu}{T^{a}}_{\mu\nu}-S^{\nu ab}\Omega_{ab,\ \mu\nu}-j^{\nu}F_{\mu\nu}\right)\right]\\
 &  & -\int d^{d}x\sqrt{\left|g\right|}\frac{1}{\sqrt{\left|g\right|}}\nabla_{\mu}\left(\sqrt{\left|g\right|}\tau_{a}^{\nu}\xi^{a}\right).
\end{eqnarray*}
Hence, the symmetry under the covariant Lie derivative implies 
\begin{equation}
\left(\nabla_{\nu}+{T^{\rho}}_{\nu\rho}\right)\tau_{a}^{\nu}-e_{a}^{\mu}\left(\tau_{a}^{\nu}{T^{a}}_{\mu\nu}-S^{\nu ab}\Omega_{ab,\ \mu\nu}-j^{\nu}F_{\mu\nu}\right)=0,
\end{equation}
which can be verified by using the equations of motion instead of
the symmetry argument here. 

We can further define the Killing vector $K^{\mu}\partial_{\mu}$
as 
\[
\delta_{K}^{C}e_{\nu}^{*a}=\delta_{K}^{C}\omega_{ab\nu}=\delta_{K}^{C}A_{\nu}=0.
\]
 Then, $\delta\mathcal{L}$ must be zero, so there is 
\begin{equation}
\frac{1}{\sqrt{\left|g\right|}}\nabla_{\mu}\left(\sqrt{\left|g\right|}\tau_{a}^{\nu}K^{a}\right)=0,\label{eq:energy_momentum_conservation}
\end{equation}
which is the energy-momentum current conservation. 
\end{widetext}

\bibliographystyle{apsrev4-1}

\end{document}